\documentclass[12pt,a4paper]{article}

\usepackage{amsfonts}
\usepackage{amsmath}
\usepackage{amsbsy}
\usepackage{theorem}
\usepackage{graphicx}
\usepackage{amssymb}
\usepackage{epstopdf}

\newtheorem{theorem}{Theorem}

\newtheorem{remark}{Remark}

\begin{document}
\sf
\title{\normalsize \bfseries SPATIAL COX PROCESSES IN AN INFINITE--DIMENSIONAL FRAMEWORK
}
\date{}
\author{\normalsize M. P. Fr\'{\i}as, A. Torres-Signes and    M. Dolores Ruiz--Medina}
\maketitle

\begin{abstract}

We introduce a new  class of spatial Cox processes driven by a Hilbert--valued  random log--intensity.  We adopt a parametric framework in the spectral domain, to estimate its spatial functional correlation structure. Specifically, we consider a spectral functional, based on the periodogram operator, inspired on Whittle estimation methodology.  Strong-consistency of the parametric estimator is proved in the linear case. We illustrate this property in a simulation study  under a  Gaussian first order Spatial Autoregressive Hilbertian scenario for the  log--intensity  model.   Our method is applied to the spatial functional prediction of respiratory  disease mortality  in the Spanish Iberian Peninsula, in the period 1980--2015.
\end{abstract}

\medskip


\section{Introduction}
\label{Intro}
Spatial point
processes  constitute an important  branch of  stochastic modeling and statistics  for  countable point sets on a planar space, generated by  a random   mechanism. These processes are applied in
many different fields such as geology, seismology, economics, image processing,
ecology, or biology.  Particularly,  the close relation between point processes and geostatistical data has been largely exploited in
 the field of spatio--temporal correlation analysis.  There exists indeed an extensive literature on statistical  modeling and analysis of point processes (see, e.g., \cite{MollerWaa04}, \cite{Daley8808}; \cite{Illian08};  \cite{Diggleb},  among others). The reader is referred to \cite{MollerWaa04} and \cite{Daley8808}, and the references therein, for a theoretical background.

The Poisson process is the most basic and simplest model of point processes. This process can be used to build
a more flexible and fundamental class of  models, named Cox processes. A Cox process (also called doubly stochastic Poisson process) is obtained as an extension of a Poisson process by considering the intensity function of a non--homogeneous Poisson process a realization of a  random  function (i.e., a stochastic process in the one--parameter case, or a random field in the multiparameter case). Cox processes are natural models for point process phenomena that are environmentally driven, but much less natural for phenomena driven primarily by interactions amongst the points  (see, e.g., \cite{MollerWaa04}). Cox processes were already introduced and studied in \cite{Cox55} (see also  \cite{Grandell76} and \cite{Stoyan95}). An attractive feature of Cox processes
is the characterization of its marginal distributions from the higher--order moments of the corresponding random intensity function.

The literature offers several subclasses of Cox processes of particular interest. We call our attention here to log-Gaussian Cox processes (LGCP) (see, e.g.,  \cite{Moller98}) which are defined as Cox processes with a random intensity being the exponential of a Gaussian stochastic process or random field. The log--normal intensity model provides a  flexible framework in spatial and spatio--temporal point pattern analysis (see \cite{Diggle}; \cite{Gonzalez16}).  The complete characterization of this process class by the  intensity and
second order   product density
 makes possible its application in different fields (see, e.g., \cite{Rathbun94} in pine forest; \cite{Serra14} in wildfire occurrences). Some extensions have also been  formulated in  \cite{Moller14}; \cite{Simpson16}; \cite{Waagepetersen16}, among others. Our paper goes beyond the real--valued case, by introducing Cox processes driven by a  spatial  Hilbert--valued ($\mathcal{H}$--valued) log--intensity random field.

 Spatio-temporally indexed data have become more widely available in many scientific fields driving an acceleration of methodological developments. This includes point patterns in space and time (see a review in \cite{Gonzalez16}), real-marked spatial point patterns or multivariate spatial point patterns (\cite{Jalilian15}; \cite{Waagepetersen16}). So far, the literature has focussed on the Euclidean plane, and develops the theoretical tools under this paradigm. In particular,  in \cite{Waagepetersen16}, a multivariate version of log-Gaussian Cox processes is proposed  to perform   statistical analysis of  multivariate point pattern data, whose cross-pair correlation
functions are given in the Euclidean space. The pair correlation function is one of most informative second order summary statistics of
 a spatial point process (\cite{MollerWaa04}; \cite{Illian08}). Kernel-based non--parametric estimation of the pair correlation function is commonly  found in the literature (\cite{Moller98}; \cite{Gonzalez16}). Some alternatives exist in a Bayesian framework (see \cite{YueLoh13}). We also refer to the  componentwise approach, based on  orthogonal series density estimators  of the pair correlation function  (see, e.g., \cite{Jalilian19}, where consistency and asymptotic normality is proved).

 Kernel estimators are computationally fast but  suffer from strong bias for spatial lags close to zero. This is a major drawback if one attempts to infer a parametric model from a non-parametric estimation,  given that the behavior near zero is important for determining the right parametric model
(see \cite{Jalilian13}).  Thus, parametric estimation should be pursued, particularly,   for first and second order moment-based analysis. However, there are not many attempts in this line. \cite{Guan08} consider pairs of spatial point processes with intensity functions sharing a common multiplicative term, and introduce a conditional likelihood estimation approach to fit a parametric model for the pair correlation function. They establish
the consistency of the resulting estimator, and discuss how the parametric estimator can be applied in model diagnostics and inference on regression parameters for the intensity functions. \cite{Waage09} propose parameter estimation for inhomogeneous spatial point processes with a regression model for the intensity function, under tractable second order properties ($K$-function). Regression parameters are estimated from a Poisson likelihood score function. In a second step, minimum contrast estimation is applied to the residual  parameters. Asymptotic normality of parameter estimators is established under certain mixing conditions. The present paper adopts  a parametric framework  in the spectral domain for the estimation of the spatial functional correlation structure. Specifically,  a  Whittle--like functional based on  the periodogram operator is considered.

   Summarizing, a certain imbalance is observed in some methodological\linebreak  branches of the statistics for point processes. However, in general terms, the literature in the field of spatial and spatio--temporal point processes has presented  a remarkable growth, in the parametric (likelihood, pseudo-li\-ke\-li\-hood,  composite likelihood), semi-parametric and non-parametric frameworks, and from classical and Bayesian perspectives (see \cite{Baddeleyeta06}; \cite{Guan06};  \cite{Diggleetal10b}; \cite{GB18},  and the references therein).

In a parallel vein, Functional Data Analysis (FDA) techniques are well suited to estimate summary statistics, which are  functional in nature. In particular, point process data classification, based on second order statistics, can be performed applying  FDA methodologies   (see, e.g., pp. 135--150 in \cite{Baddeleyeta06}, and \cite{Illian08}). But FDA is a relatively new branch in point pattern analysis.
  A functional approach is proposed in \cite{WMZ} to obtain the covariance structure
of the  random densities, in the case where the shapes of the intensity functions that generate the
observed event times are not known. A reconstruction formula is derived for the  object-specific density functions to approximate  the distribution  of event times observed over a fixed
time interval. Another attempt, that contributes to the infinite--dimensional counting process  framework can be found in \cite{Bosq14}, where  an $\ell^{2}$-valued homogeneous Poisson process is introduced, addressing parameter estimation and prediction, from  classical and Bayesian frameworks. The    equivalent asymptotic efficient behavior of both approaches is also proved. In \cite{Torres16}, a family of $\ell^{2}$-valued temporal log-Gaussian Cox processes  is introduced. An Autoregressive Hilbertian  (AR$\mathcal{H}$(1)) process--based estimation methodology  is  applied for functional prediction of the     Gaussian log--intensity.

This paper introduces a new class of spatial Cox processes, driven by a Hilbert--valued random log--intensity. We consider  the weak--sense definition of the
functional values of the log--intensity. We refer the resulting generalized log--intensity random field model as the generalized log--risk process.   The conditional marginal probability distributions of the associated counting random measure are Poisson distributed.  As usual, in the log--Gaussian case,   the spatial functional correlation structure of  the random  log--intensity  characterizes  higher  order moments.  Under stationarity in space,   a parametric framework is adopted here in the spatial functional spectral domain. In the spirit of Whittle estimation methodology,  second order summary functional statistics are then  approximated  in terms of a   spectral functional, based on the periodogram and  spectral density operators. Our loss function measures the proximity of the empirical spatial functional correlation structure of the data, reflected by the periodogram operator, and the elements of the parametric spectral density operator family tested.     Strong-consistency of the formulated parametric estimator of the  second order  product density is proved.

To illustrate our results,   a simulation study is undertaken under a Spatial Autoregressive Hilbertian (SAR$\mathcal{H}$(1))   process  framework (see \cite{Ruiz11a}).  We apply our approach  to the parametric estimation of the eigenvalues and eigenvectors of the involved  autocorrelation operators.  These parametric estimators allow the approximation of the spectral density operator.  The strong and mean--square consistency  are illustrated.   Particularly, in the first numerical example,  we estimate  the hyperparameter defining the support of the eigenvectors of the involved autocorrelation operators.   The problem of estimating scale and localization  hyperparameters, characterizing the eigenvalues of the autocorrelation operators, is  addressed in a second numerical example.   Our estimation methodology is  also implemented with  real--data. Namely, we analyze respiratory disease mortality in the Spanish Iberian Peninsula during  the period 1980--2015. The Spanish National Statistical Institute  provided the  data, consisting of 432 monthly records on respiratory disease mortality at the 48  Spanish  provinces.

The outline of the paper is the following. Section \ref{s3b} introduces the  spatial Cox   process  class,  driven by a spatial  Hilbert--valued    log--intensity. Section \ref{pepo} derives the strong--consistent parametric estimation of its  spatial functional correlation structure in the spectral domain.  A simulation study illustrates  in Section \ref{simulationstdy} the properties of the presented spatial functional estimation approach. This  approach is validated using real data in Section  \ref{s6}.    The paper ends with some final discussion.

\section{Cox processes driven by a spatial $\mathcal{H}$--valued  random log--intensity}
\label{s3b}
In what follows,  denote by  $(\Omega,\mathcal{A},P)$  the probability space over which we define all random variables in this paper.
In the subsequent developments, we  omit the dependency on $\omega \in \Omega$ of  all measurable mappings, from   $(\Omega,\mathcal{A},P)$ to the real or complex line, as well as  from   $(\Omega,\mathcal{A},P)$ to a function space. We only reflect such  dependency the first time, when  conditional probability distributions are introduced from the observation of their sample values. Usually, the subindex denotes the location of the element of a  given family of operators or  random variables, and the argument refers to the element located on the set defining its support.

 Let $\mathcal{H}$  be   a real separable Hilbert space   of functions, and denote by $\mathcal{H}+i\mathcal{H}$    its complex version. In the following,  $\left\langle \cdot,\cdot\right\rangle$ and  $\|\cdot\|$ respectively  denote the inner product and norm  on the space  $\mathcal{H}+i\mathcal{H}.$ The same notation will be used for the inner product and norm of $\mathcal{H},$ considered as a  subspace of $\mathcal{H}+i\mathcal{H}.$
For practical purposes we refer to $\mathcal{H}=L^{2}(\mathcal{T}),$ the space of square--integrable functions on the time interval $\mathcal{T}.$   In this section, $B$ will stand for a bounded Borel set $B\in \mathcal{B}^{d}.$

 Denote by $\left\{\varkappa_{\mathbf{z}}, \ \mathbf{z}\in \mathbb{R}^{d}\right\}$  a spatial random  process with values in $\mathcal{H}.$  Applying  Riesz Representation Theorem,  we can  define a random functional $ X_{\mathbf{z}} $  on $ \mathcal{H} $ by  the identity
\[
X_{\mathbf{z}} (\varphi) = \left\langle \varkappa_{\mathbf{z}},\varphi\right\rangle,\quad  \forall \varphi \in\mathcal{H}, \quad  \mathbf{z}\in \mathbb{R}^{d}.
\]
  From a technical point of view, this definition will help us to introduce different concepts related to the probability distribution, and spatial extrapolation of $\mathcal{H}$--valued random variables.
  \begin{remark}
\label{rem1ir}  Note that under the assumption
  \begin{equation}
\sum_{p=1}^{\infty} E\left|\left\langle \varkappa_{\mathbf{z}},\phi_{p}\right\rangle\right|^{2}=\sum_{p=1}^{\infty}E\left| X_{\mathbf{z}}(\phi_{p})\right|^{2}<\infty,\quad \mathbf{z}\in \mathbb{R}^{d},\label{condhvalued}\end{equation}
\noindent for any  orthonormal basis $\{\phi_{p},\ p\geq 1\}$ of $\mathcal{H},$
 $\varkappa_{\mathbf{z}}$ is a random element in $\mathcal{H},$ i.e.,   $P[\varkappa_{\mathbf{z}}\in \mathcal{H}]=1,$ for any $\mathbf{z}\in \mathbb{R}^{d}$
 (see \cite{LedouxT91}).   Hence, both identities,  strong--sense (pointwise), and  weak--sense (as a functional on $\mathcal{H}$) identities  will appear throughout this paper. For example,
 in equation  (\ref{intp}) below the pointwise definition of $\varkappa_{\mathbf{z}}\in \mathcal{H}$ is considered, while  in equation (\ref{asi})   its weak--sense definition is applied.

 \end{remark}

   Consider  now the spatial functional  random intensity $\boldsymbol{\Lambda }=\{\Lambda_{\mathbf{z}}(\cdot ), \ \mathbf{z}\in \mathbb{R}^{d}\}$ defined as
\begin{equation}\Lambda_{\mathbf{z}}(t)=
\exp\left(\varkappa_{\mathbf{z}}(t)\right)=\sum_{k=1}^{\infty}\frac{\left[\varkappa_{\mathbf{z}}(t)\right]^{k}}{k!},\quad \forall t\in \mathcal{T},\quad \mathbf{z}\in \mathbb{R}^{d},\label{intp}
\end{equation}
\noindent where $\varkappa_{\mathbf{z}}(t)$ denotes the pointwise value of the mapping $t\to \varkappa_{\mathbf{z}}(t),$ $t\in \mathcal{T},$
 since $\varkappa_{\mathbf{z}}\in \mathcal{H}=L^{2}(\mathcal{T}),$ almost surely.    Note that $\varkappa_{\mathbf{z}}:(\Omega,\mathcal{A},P)\to \mathcal{H}$ defines a measurable function, for any $\mathbf{z}\in \mathbb{R}^{d}.$
 By construction,   for a given $\omega \in \Omega,$ set  $\ln\left(\lambda_{\mathbf{z}}\right)(t):=\varkappa_{\mathbf{z}}(\omega ,t)=\mathcal{X}_{\mathbf{z}}(t),$  for every $t\in \mathcal{T},$ and  $\mathbf{z}\in\mathbb{R}^{d}.$  The    realizations  $\boldsymbol{\lambda }=\{\lambda_{\mathbf{z}}(\cdot), \ \mathbf{z}\in \mathbb{R}^{d}\}$   of  $\boldsymbol{\Lambda }=\{\Lambda_{\mathbf{z}}(\cdot ),$  $\mathbf{z}\in \mathbb{R}^{d}\}$ are  then  almost surely (a.s.)   positive. As commented before,  we omit the dependence on $\omega \in \Omega$ of the realizations $\left\{\varkappa_{\mathbf{z}}(\omega ,\cdot)=\mathcal{X}_{\mathbf{z}}(\cdot),\ \mathbf{z}\in \mathbb{R}^{d},\ \omega \in \Omega \right\}$  of the spatial $\mathcal{H}$--valued log--intensity $\{\varkappa_{\mathbf{z}},\ \mathbf{z}\in \mathbb{R}^{d}\}.$

 The following condition is now considered.

\medskip

\noindent \textbf{Assumption A1}. For  any $B\in \mathcal{B}^{d},$ assume
\begin{equation}
\int_{B}\exp\left(X_{\mathbf{z}}(\varphi )\right)d\mathbf{z}=\int_{B}\exp\left(\left\langle \varkappa_{\mathbf{z}},\varphi \right\rangle\right)d\mathbf{z}<\infty,\quad \forall \varphi \in \mathcal{H},
\label{asi}
\end{equation}
\noindent in the norm  of the space  $\mathcal{L}^{2}(\Omega,\mathcal{A},P),$ given by $\|X\|_{\mathcal{L}^{2}(\Omega,\mathcal{A},P)}^{2}=E[X^{2}],$ for any zero--mean second order random variable $X$ on $(\Omega,\mathcal{A},P).$

\begin{remark}\label{remA1}
From \textbf{Assumption A1}, the marginal probability distributions of the  introduced  Cox process family have finite  second order moments.
Equation (\ref{asi}) holds if the  process  $\left\{\varkappa_{\mathbf{z}}(t),\ t\in \mathcal{T},\ \mathbf{z}\in \mathbb{R}^{d}\right\}$  is continuous in the mean--square sense with respect to both arguments $t\in \mathcal{T},$ and $\mathbf{z}\in \mathbb{R}^{d},$ which is equivalent to assume that the kernel  $r_{\mathbf{z},\mathbf{y}}(t,s)=E[\varkappa_{\mathbf{z}}(t)\varkappa_{\mathbf{y}}(s)],$     defining the covariance operator $\mathcal{R}_{\mathbf{z},\mathbf{y}}=E[\varkappa_{\mathbf{z}}\otimes \varkappa_{\mathbf{y}}]$ is continuous in  $t,s\in  \mathcal{T},$  and  $\mathbf{z},\mathbf{y}\in \mathbb{R}^{d}.$ Here, $\varkappa_{\mathbf{z}}\otimes \varkappa_{\mathbf{y}}$ denotes the tensorial product of two random elements  in $\mathcal{H}.$  Hence, it defines a Hilbert-Schmidt operator on $\mathcal{H}$ almost surely.
\end{remark}

Let now $N:(\Omega ,\mathcal{A},\mathcal{P})\times \mathcal{B}^{d}\to \mathbb{N}$ be a random counting measure over the Borel $\sigma$--algebra $\mathcal{B}^{d}$ of $\mathbb{R}^{d}.$  That is, for each $\omega \in \Omega $ fixed,  $N(\omega , B),$ $B\in \mathcal{B}^{d},$ defines a counting measure  on the sets of $\mathcal{B}^{d},$ while, for each set $B\in \mathcal{B}^{d},$ $N(\omega ,B),$ $\omega \in \Omega ,$ defines a integer--valued random variable on $(\Omega,\mathcal{A},P).$    We assume that, for any  $\varphi \in \mathcal{H},$
   the conditional probability distribution of the number  $N(B)$ of random  events     that occur on a set  $B,$
   given the spatial realization $x(\varphi)=\{x_{\mathbf{z}}(\varphi)=\left\langle \mathcal{X}_{\mathbf{z}},\varphi\right\rangle, \ \mathbf{z}\in \mathbb{R}^{d}\}$  of $X(\varphi),$
   follows a Poisson probability distribution with  mean $\int_{B}\exp\left(x_{\mathbf{z}}(\varphi )\right)d\mathbf{z},$
 for any  $B\in \mathcal{B}^{d}.$ Here, as before, for every $\varphi \in \mathcal{H},$ $x(\varphi)=\{x_{\mathbf{z}}(\varphi)=\left\langle \mathcal{X}_{\mathbf{z}},\varphi\right\rangle, \ \mathbf{z}\in \mathbb{R}^{d}\}$ is computed from  the  sample curves   $\mathcal{X}_{\mathbf{z}}\in \mathcal{H},$ spatially distributed over $\mathbf{z}\in \mathbb{R}^{d}.$

\begin{remark} \label{rem2rrth}

We consider the  terminology  \emph{test function} in a wide sense from Riesz Representation Theorem. Note that FDA techniques are usually applied after   interpolation and smoothing of the data. Local singular behaviors are then regularized. Alternatively,
 when high--singular systems are analyzed, the space $C_{0}^{\infty}(\mathcal{T})$ of infinitely differentiable functions with compact support in $\mathcal{T}$ is usually considered as test function space, allowing the regularization of the system.   \end{remark}

 \subsection{Least--squares prediction  and marginal   second order moments}
 We introduce the first and second order  moments of the   marginal  conditional  probability distributions of our random measure $\{N(B); \ |B|<\infty, \ B\in \mathcal{B}^{d}\},$ where $|B|=\int_{B}d\mathbf{z}.$

For $B\in \mathcal{B}^{d}$ and $\varphi\in \mathcal{H},$ the conditional Poisson probability distribution of $N(B),$  given $x_{\mathbf{z}}(\varphi),$  $\mathbf{z}\in \mathbb{R}^{d},$   leads to
  \begin{eqnarray}
  & &f_{N(B)/x_{\mathbf{z}}(\varphi),  \mathbf{z}\in \mathbb{R}^{d}}(t):= E\left[\exp\left(tN(B) \right)/ x_{\mathbf{z}}(\varphi),  \mathbf{z}\in \mathbb{R}^{d}\right]\nonumber\\ && \hspace*{0.5cm}=
   \exp\left(\left[\int_{B}\exp\left(x _{\mathbf{z}}(\varphi )\right)d\mathbf{z}\right]\left(\exp(t)-1\right)\right),\quad  t \in \mathbb{R},\quad \forall \varphi\in \mathcal{H}.
   \label{cm}
\end{eqnarray}
 As usual, the least--squares predictor $\widehat{N(B)}$ of $N(B)$ is  computed from the first derivative of   (\ref{cm}) at $t=0.$    Thus,
\begin{equation}\widehat{N(B)}=
E\left[N(B)/x_{\mathbf{z}}(\varphi ),\ \mathbf{z}\in \mathbb{R}^{d}\right]=\int_{B}\exp\left(x_{\mathbf{z}}(\varphi )\right)d\mathbf{z},
\label{lspredictor}
\end{equation}

\noindent which  coincides with  the conditional variance.   The corresponding variance decomposition formula is computed from the following identities, obtained from the second derivative of
(\ref{cm}) at $t=0$
\begin{eqnarray}
E\left[\mbox{Var}\left(N(B)/X_{\mathbf{z}}(\varphi ),\ \mathbf{z}\in \mathbb{R}^{d} \right)\right] &=&
\int_{B}E\left[\exp\left(X_{\mathbf{z}}(\varphi)\right)\right]d\mathbf{z}\nonumber\\
\mbox{Var}\left(E\left[N(B)/X_{\mathbf{z}}(\varphi),\ \mathbf{z}\in \mathbb{R}^{d}\right]\right)&=&
\int_{B\times B}E\left[\exp\left(X_{\mathbf{z}}(\varphi )+X_{\mathbf{y}}(\varphi )\right)\right]d\mathbf{z}d\mathbf{y}\nonumber\\
&-& \int_{B\times B}E\left[\exp\left(X_{\mathbf{z}}(\varphi )\right)\right]E\left[\exp\left(X_{\mathbf{y}}(\varphi )\right)\right]d\mathbf{z}d\mathbf{y},
\nonumber\\\label{dvf}
\end{eqnarray}

\noindent where from (\ref{lspredictor})
  \begin{eqnarray}&&\hspace*{-0.5cm} E_{\varphi}\left[N(B)\right]=
\int_{B} E\left[\exp\left(X_{\mathbf{z}}(\varphi )\right)\right]d\mathbf{z},\label{lspredictor30}\end{eqnarray}
\noindent  for $B\in \mathcal{B}^{d}$ and for any $\varphi\in \mathcal{H}.$
\noindent Thus, from (\ref{dvf}) and  (\ref{lspredictor30}), we obtain
\begin{eqnarray}
&&\mbox{Var}_{\varphi}\left(N(B)\right)=\int_{B}E\left[\exp\left(X_{\mathbf{z}}(\varphi )\right)\right]d\mathbf{z}\nonumber\\
& &\hspace*{1.3cm}+\int_{B\times B}E\left[\exp\left(X_{\mathbf{y}}(\varphi )+X_{\mathbf{z}}(\varphi )\right)\right]d\mathbf{y}d\mathbf{z}\nonumber\\
&&\hspace*{1.3cm}-\int_{B\times B}E\left[\exp\left(X_{\mathbf{y}}(\varphi )\right)\right]E\left[\exp\left(X_{\mathbf{z}}(\varphi )\right)\right]d\mathbf{y}d\mathbf{z}.
\label{lspredictor3}
\end{eqnarray}

In the spatial stationary log--Gaussian case,  equations  (\ref{lspredictor30})--(\ref{lspredictor3}) can be rewritten   in terms of the second order moments of a zero--mean Gaussian spatial $\mathcal{H}$--valued  log-intensity  $\left\{\varkappa_{\mathbf{z}},\ \mathbf{z}\in \mathbb{R}^{d}\right\}.$
 For every $\varphi\in \mathcal{H}$ and set $B\in \mathcal{B}^{d},$
 \begin{eqnarray}E_{\varphi}\left[N(B)\right]&=&
\exp\left(\frac{\mathcal{R}_{0}(\varphi)(\varphi)}{2}\right)|B|\nonumber\\
\mbox{Var}_{\varphi}\left(N(B)\right)
  &=& \exp\left(\mathcal{R}_{0}(\varphi)(\varphi)\right)\int_{B\times B}  \exp\left(\frac{\mathcal{R}_{\mathbf{z}-\mathbf{y}}(\varphi)(\varphi)+\mathcal{R}_{\mathbf{y}-\mathbf{z}}(\varphi)(\varphi)}{2}\right)  d\mathbf{z}d\mathbf{y}
  \nonumber\\
&&+|B|\exp\left(\frac{\mathcal{R}_{0}(\varphi)(\varphi)}{2}\right)\left[1-|B|\exp\left(\frac{\mathcal{R}_{0}(\varphi)(\varphi)}{2}\right)\right],
 \label{kernelsncm2}\end{eqnarray}
\noindent where, as before,  $|B|$ denotes the Lebesgue measure of  $B,$ and $\mathcal{R}_{\mathbf{z}-\mathbf{y}}$ is the spatial covariance operator of $\{\varkappa_{\mathbf{z}},\ \mathbf{z}\in \mathbb{R}^{d}\}$  given by

\begin{eqnarray}
\mathcal{R}_{\mathbf{z}-\mathbf{y}}(f)(g)&=&
E\left\langle X_{\mathbf{z}}\otimes X_{\mathbf{y}}(f), g
\right\rangle =E\left[ X_{\mathbf{z}}(g)X_{\mathbf{y}}(f)\right]
\nonumber\\
&=& E\left[\left\langle \varkappa_{\mathbf{z}},g \right\rangle\left\langle \varkappa_{\mathbf{y}},f\right\rangle\right],\ \forall f,g\in \mathcal{H},\quad \mathbf{z}, \mathbf{y}\in \mathbb{R}^{d}.
\label{SCO}
\end{eqnarray}

The $n$-order product density is then  computed from the intensity and second order product density as follows. For $\varphi\in \mathcal{H},$

\begin{eqnarray}
&&\rho^{(n)}_{\varphi}(\mathbf{z}_{1},\dots, \mathbf{z}_{n})= E\left[\prod_{i=1}^{n}\exp\left(x_{\mathbf{z}_{i}}(\varphi )\right)\right]=E\left[\exp\left(\sum_{i=1}^{n}x_{\mathbf{z}_{i}}(\varphi )\right)\right]\nonumber\\
&&=[\rho_{\varphi}]^{n}\exp\left(\frac{1}{2}\sum_{i=1}^{n}\sum_{j=1}^{n}\mathcal{R}_{\mathbf{z}_{i}-\mathbf{z}_{j}}
(\varphi)(\varphi)\right),\quad  \mathbf{z}_{1},\dots, \mathbf{z}_{n}\in \mathbb{R}^{d},
  \label{kernelsncm3}\end{eqnarray}
\noindent with the spatial functional intensity $\rho_{\varphi}$ defined  as
\begin{eqnarray}
\rho_{\varphi } =\rho^{(1)}_{\varphi }(\mathbf{z})&=&\exp\left(\frac{\mathcal{R}_{0}(\varphi)(\varphi)}{2}\right),\quad \forall \mathbf{z}\in \mathbb{R}^{d},\quad \varphi\in \mathcal{H}.
\label{eqfoif}
\end{eqnarray}

\noindent In the isotropic case,  for $i,j=1,\dots,n,$   the spatial  pair correlation functional is defined from the identity  \begin{eqnarray}
g_{\varphi}(\mathbf{z}_{j}-\mathbf{z}_{i}) &=& \frac{\rho^{(2)}_{\varphi}(\mathbf{z}_{i}-\mathbf{z}_{j})}{[\rho_{\varphi }]^{2}}=\exp\left(\mathcal{R}_{\mathbf{z}_{i}-\mathbf{z}_{j}}
(\varphi)(\varphi)\right),\quad \forall  \varphi\in \mathcal{H}.\label{eqsoif}
\end{eqnarray}

\section{Strong--consistent estimation of the spatial  functional correlation structure of the log--intensity}
\label{pepo} We propose a strong--consistent parametric estimator of the spectral density operator, inspired on the Whittle  functional (see \cite{Panaretos13} in the nonparametric framework).
We restrict our attention here to the case $\{\varkappa_{\mathbf{z}}, \ \mathbf{z}\in \mathbb{Z}^{d}\},$ i.e.,  the $\mathcal{H}$--valued  log--intensity is observed on a  spatial regular grid. We then consider the interval  $[-\pi,\pi]^{d}$ as the index set of our  spectral density operator family $\{\mathcal{F}_{\boldsymbol{\xi }},\ \boldsymbol{\xi }\in [-\pi,\pi]^{d}\},$ given by
 \begin{equation}
 \mathcal{F}_{\boldsymbol{\xi }}(h)(g)=\frac{1}{(2\pi)^{d}}\sum_{\mathbf{z}\in \mathbb{Z}^{d}}\exp\left(-i\sum_{j=1}^{d}\xi_{j}z_{j}\right)\mathcal{R}_{\mathbf{z}}(h)(g),\quad \boldsymbol{\xi }\in [-\pi,\pi]^{d},\label{sdo}
 \end{equation}

\noindent  for all $h,g\in \mathcal{H}+i\mathcal{H}.$ We assume the series $\frac{1}{(2\pi)^{d}}\sum_{\mathbf{z}\in \mathbb{Z}^{d}}\exp\left(-i\sum_{j=1}^{d}\xi_{j}z_{j}\right)\mathcal{R}_{\mathbf{z}}$ converges in the trace norm  (see, e.g.,  \cite{Panaretos13}), where, for any $\mathbf{z}\in \mathbb{R}^{d},$ the covariance operator $\mathcal{R}_{\mathbf{z}}$ has been introduced in  (\ref{SCO}).

We assume the following condition.

\medskip

\noindent \textbf{Assumption A2}. Let $\mathbf{z}=(z_{1},\dots,z_{d})\in \mathbb{Z}^{d},$  with    $z_{i}\in [-\mathcal{N}_{i}+1,\mathcal{N}_{i}-1],$ $i=1,\dots,d,$   and   $\mathcal{N}=\mathcal{N}_{1}\times\dots\times \mathcal{N}_{d}.$ For an orthonormal basis  $\{\psi_{k},\ k\geq 1\}$
of $\mathcal{H},$ assume that as $\mathcal{N}\to \infty,$
$$\sum_{k,l\geq 1}\left|\left[\frac{1}{\mathcal{N}}\left[\sum_{y_{1}=1}^{\mathcal{N}_{1}-z_{1}}\dots \sum_{y_{d}=1}^{\mathcal{N}_{d}-z_{d}} X_{\mathbf{y}}(\psi_{k})X_{\mathbf{y}+\mathbf{z}}(\psi_{l})\right]-\mathcal{R}_{\mathbf{z}}(\psi_{k})(\psi_{l})\right]\right|^{2}\to_{\mbox{a.s.}} 0,$$
\noindent where we have used the notation $\mathbf{y}=(y_{1},\dots,y_{d})\in \prod_{i=1}^{d}[1,\mathcal{N}_{i}-z_{i}]\cap \mathbb{Z}^{d}.$

 This condition provides the ergodicity, in the second order moment sense, with respect to the Hilbert--Schmidt operator norm $\|\cdot\|_{\mathcal{S}(\mathcal{H})}$  of the spatial functional log--intensity process  $\left\{\varkappa_{\mathbf{z}},\ \mathbf{z}\in \mathbb{Z}^{d}\right\}.$

   \begin{remark} As usual, strong--consistency in the $\mathcal{H}$--valued context  is defined from the a.s. convergence in the norm $\|\cdot\|_{\mathcal{H}}.$ This fact is applied in \textbf{Assumption A2}, considering the Hilbert space $\mathcal{S}(\mathcal{H})$ of Hilbert--Schmidt operators on $\mathcal{H}.$ Thus
\textbf{Assumption A2} defines the strong--consistency of the  empirical  functional  second order  moments of $\varkappa_{\mathbf{z}}$ with respect to $\mathcal{S}(\mathcal{H})$--norm.
  \end{remark}

  \begin{remark} \textbf{Assumption A2} is satisfied, in particular, when $$\sum_{\mathbf{z}\in \mathbb{R}^{d}}  \|\mathcal{R}_{\mathbf{z}}\|_{L^{1}(\mathcal{H})}<\infty,$$  \noindent with, for $\mathbf{z}\in \mathbb{R}^{d},$ $\|\mathcal{R}_{\mathbf{z}}\|_{L^{1}(\mathcal{H})}=\sum_{k\geq 1}\left\langle [\mathcal{R}_{\mathbf{z}}^{\star}\mathcal{R}_{\mathbf{z}}]^{1/2}(\varphi_{k}),\varphi_{k}\right\rangle,$ for any orthonormal basis $\{\varphi_{k},\ k\geq 1\}$ of $\mathcal{H}$ (see  Corollary 2.3 in \cite{Bosq2000}). Under this condition, the spectral density operator $\mathcal{F}_{\boldsymbol{\xi}}$ is a nuclear operator for any $\boldsymbol{\xi}\in [-\pi,\pi]^{d}.$

  \end{remark}

Consider now a     parametric family of
spectral density operators $$\left\{\mathcal{F}_{\boldsymbol{\xi} ,\theta},\ \boldsymbol{\xi} \in
[-\pi,\pi]^{d}, \theta \in \Theta \subset
\mathbb{R}^{q}\right\},$$\noindent  where
the compact parameter space  $\Theta \subset \mathbb{R}^{q},$ $q\geq 1,$ is such that  \begin{eqnarray}&&\left\|\int_{[-\pi,\pi]^{d}}\mathcal{F}_{\boldsymbol{\xi} ,\theta_{1}}\mathcal{F}_{\boldsymbol{\xi},\theta_{2}}^{-1}d\boldsymbol{\xi} \right\|_{\mathcal{L}(\mathcal{H}+i\mathcal{H})}=\left\|\int_{[-\pi,\pi]^{d}}\mathcal{F}_{\boldsymbol{\xi} ,\theta_{2}}^{-1}\mathcal{F}_{\boldsymbol{\xi },\theta_{1}}d\boldsymbol{\xi}
\right\|_{\mathcal{L}(\mathcal{H}+i\mathcal{H})}< \infty,\nonumber\\ \label{cbn}
\end{eqnarray}
\noindent for every $\theta_{1},\theta_{2}\in \Theta.$ Here, $\|\cdot\|_{\mathcal{L}(\mathcal{H}+i\mathcal{H})}$ denotes the norm in the space of bounded linear operators on $\mathcal{H}+i\mathcal{H}.$

Note that Examples 1 and  2 in Section \ref{simulationstdy} provide a suitable scenario where  condition  (\ref{cbn}) is satisfied in the functional linear setting. Specifically,  in the one--parameter scenario of Example 1,  $\theta $ characterizes  the support of the  eigenvectors of the autocorrelation operators. In Example 2, the  components of $\theta $ are location and scale parameters involved in the definition of  the parameterized eigenvalues of the autocorrelation operators.  Particularly, the presented approach extends to the infinite--dimensional framework the  parametric linear modeling   adopted in  \cite{Hannan73} in the real--valued case.

 The following  four assumptions are  considered in the  functional linear parametric framework adopted in  Theorem 1 below. For an  arbitrary orthonormal basis $\{\varphi_{k},\ k\geq 1\}$  of $\mathcal{H}+i\mathcal{H}$   assume:

\noindent (C1) For $\theta ,\theta^{\prime}\in \Theta,$  with
$\theta \neq \theta^{\prime},$ the set $\{\boldsymbol{\xi}\in [-\pi,\pi]^{d};\
\mathcal{F}_{\boldsymbol{\xi} ,\theta}\neq \mathcal{F}_{\boldsymbol{\xi} ,\theta^{\prime
}} \}$ has positive Lebesgue measure. Hence, different $\theta$
values correspond to different spatial functional dependence structures.

\noindent (C2) Operator $\mathcal{F}_{\boldsymbol{\xi} ,\theta}$ is continuous over $\boldsymbol{\xi}\in [-\pi,\pi]^{d},$ and  $\theta\in \Theta ,$ with respect to $\mathcal{L}(\mathcal{H}+i\mathcal{H})$--norm. That is, $\left\|\mathcal{F}_{\boldsymbol{\xi},\theta }-\mathcal{F}_{\boldsymbol{\xi}^{\star },\theta^{\star }}\right\|_{\mathcal{L}(\mathcal{H}+i\mathcal{H})}\to 0,\quad (\boldsymbol{\xi} ,
\theta )\to (\boldsymbol{\xi}^{\star }, \theta^{\star}),$ \linebreak for all $(\boldsymbol{\xi}^{\star },\theta^{\star})\in [-\pi,\pi]^{d}\times \Theta .$

\noindent (C3)  The  spatial functional infinite--order moving average representation of  the log--intensity process is given in terms of a  generalized $\mathcal{H}$--valued white noise innovation process having identity autocovariance operator. Thus,
\begin{equation}\int_{[-\pi,\pi]^{d}}\log\left((2\pi)^{d}\mathcal{F}_{\boldsymbol{\xi} ,\theta
}(\varphi_{k})(\varphi_{k})\right)d\boldsymbol{\xi}=0,\quad \forall k\geq 1,\ \theta \in \Theta.\label{sn}\end{equation}

 \noindent (C4) The Cesaro sum of the Fourier series of $\mathcal{F}_{\boldsymbol{\xi} ,\theta}^{-1}(\varphi_{k}) (\varphi_{k})$
  converges uniformly in $\boldsymbol{\xi} \in [-\pi,\pi]^{d},$  $k\geq 1,$ and $\theta\in \Theta .$

\begin{remark}
\label{rem4pf} In the functional linear time series setting, the $\mathcal{H}$--valued innovation process  has usually trace autocovariance operator (see, e.g., \cite{Bosq2000}). Under (C3), we have introduced a more general framework for such innovation process,  defined by a  generalized random field model with identity autocovariance operator.
\end{remark}
\begin{remark}
\label{uospv} As pointed out in Remark \ref{rem4pf}, the  projections of the innovation process onto  an orthonormal basis  of $\mathcal{H}$ define a spatial white noise process  with unit variance. This local singular behavior displayed by our generalized innovation random field model is   regularized by convolution with a suitable  linear operator sequence. Thus, the spatial summability of the square of the Hilbert--Schmidt operator norm of this  sequence should be  assumed. This condition is equivalent to the integrability in the frequency domain of the trace norm of the spectral density operator.
\end{remark}

\medskip

 Let  $\widetilde{\varkappa}_{\boldsymbol{\xi} }^{\mathcal{N}}$ be the  functional discrete Fourier transform of the data, given by the following
 identity in the norm of $\mathcal{H}+i\mathcal{H},$

\begin{equation}\widetilde{\varkappa}_{\boldsymbol{\xi} }^{\mathcal{N}}=
\frac{1}{\sqrt{\mathcal{N}(2\pi)^{d}}}\sum_{y_{1}=1}^{\mathcal{N}_{1}}\dots\sum_{y_{d}=1}^{\mathcal{N}_{d}}
\exp\left(-i\sum_{j=1}^{d}\xi_{j}y_{j}\right)\varkappa_{\mathbf{y}},\ \boldsymbol{\xi }\in [-\pi,\pi]^{d},\ \mathcal{N}=\prod_{i=1}^{d}\mathcal{N}_{i},\label{fdft}
\end{equation}
\noindent where we have applied that $\widetilde{\varkappa}_{\boldsymbol{\xi} }^{\mathcal{N}}\in \mathcal{H}+i\mathcal{H},$ almost surely, for any $\boldsymbol{\xi}\in [-\pi,\pi]^{d},$ in view of the trace property   of the autocovariance operator of $\left\{\varkappa_{\mathbf{z}},\ \mathbf{z}\in \mathbb{R}^{d}\right\}$ (see equation (\ref{condhvalued}) in Remark  \ref{rem1ir}).
The periodogram operator is then defined as the empirical operator
\begin{eqnarray}
\mathcal{I}_{\boldsymbol{\xi} }^{\mathcal{N}}(g)(h)&=&\frac{1}{\mathcal{N}(2\pi^{d})}
\sum_{y_{1}=1}^{\mathcal{N}_{1}}\dots\sum_{y_{d}=1}^{\mathcal{N}_{d}}\sum_{z_{1}=1}^{\mathcal{N}_{1}}\dots\sum_{z_{d}=1}^{\mathcal{N}_{d}}X_{y_{1},\dots, y_{d}}^{\mathcal{N}}(h)\nonumber\\
\\ &&\hspace*{-0.3cm}\times X_{z_{1},\dots,z_{d}}^{\mathcal{N}}(g)\exp\left(-i\sum_{j=1}^{d}\xi_{j} (y_{j}-z_{j})\right),
\nonumber\end{eqnarray}

\noindent for $g,h\in \mathcal{H}+i\mathcal{H},$ and $\boldsymbol{\xi}\in [-\pi,\pi]^{d}.$  Note that from (\ref{fdft}), the trace property of  $\mathcal{I}_{\boldsymbol{\xi} }^{\mathcal{N}}$  follows straightforward from Parseval identity in  $\mathcal{H}+i\mathcal{H},$ keeping in mind that $\mathcal{I}_{\boldsymbol{\xi} }^{\mathcal{N}}$ is nonnegative.  Under \textbf{Assumption A2}, and the above formulated conditions, the following parameter estimator is considered, based on a  functional sample size $\mathcal{N},$
\begin{eqnarray}
&&\widehat{\theta}_{\mathcal{N}}=\mbox{arg min}_{\theta \in \Theta}\left[\sigma_{\mathcal{N}}(\theta )\right],
\label{we2bb}
\end{eqnarray}
\noindent  where  $\sigma_{\mathcal{N}}(\theta)$ denotes the empirical loss function
 \begin{eqnarray}  \sigma_{\mathcal{N}}(\theta)&=&
 \sup_{k\geq 1}\frac{1}{(2\pi)^{d}}\int_{[-\pi, \pi]^{d}}\left|\mathcal{F}_{\boldsymbol{\xi } ,\theta }^{-1}\mathcal{I}_{\boldsymbol{\xi} }^{\mathcal{N}}(\varphi_{k}) (\varphi_{k})\right|d\boldsymbol{\xi }\nonumber\\
 &=&  \sup_{k\geq 1}\frac{1}{(2\pi)^{d}}\int_{[-\pi, \pi]^{d}}\left|\mathcal{I}_{\boldsymbol{\xi }}^{\mathcal{N}}\mathcal{F}_{\boldsymbol{\xi } ,\theta }^{-1}(\varphi_{k}) (\varphi_{k})\right|d\boldsymbol{\xi },\quad \forall \theta \in \Theta, \label{wfs}
 \end{eqnarray}
\noindent  whose asymptotic properties will be analyzed in Theorem \ref{th1ex} below.

We   consider the corresponding parametric estimator of the spatial covariance operator
 \begin{eqnarray}&&
\hspace*{-0.7cm}\widehat{\mathcal{R}}_{\mathbf{z},\theta }(h)(g)=\frac{1}{(2\pi)^{d}}\int_{[-\pi ,\pi]^{d}}\exp\left(i\sum_{j=1}^{d} z_{j}\xi_{j}\right)  \mathcal{F}_{\boldsymbol{\xi},\widehat{\theta }_{\mathcal{N}} }(h)(g)d\boldsymbol{\xi }
\label{estscs}
\end{eqnarray}
\noindent   for $h,g\in \mathcal{H},$ and  $\mathbf{z}\in \mathbb{Z}^{d}.$ Also,  in the isotropic log-Gaussian case,   the following parametric estimator of  the pair correlation function is obtained
 \begin{eqnarray}&&
\widehat{g_{\varphi,\theta}}(\mathbf{z}) = \exp\left( \widehat{\mathcal{R}}_{\mathbf{z},\theta }(\varphi)(\varphi)\right)=  \exp\left(\mathcal{R}_{\mathbf{z},\widehat{\theta }_{\mathcal{N}}}
(\varphi)(\varphi)\right),\ \forall \varphi \in \mathcal{H},\ \mathbf{z}\in \mathbb{R}^{d}.\nonumber
\end{eqnarray}
\noindent In this line of thinking, and though the next result is derived beyond the log--Gaussian assumption, recall that, under the log--Gaussian Cox scenario, the strong--consistency proved in  Theorem \ref{th1ex} also holds for the parametric estimator of the product density \begin{eqnarray}\rho^{(n)}_{\varphi,\widehat{\theta }_{\mathcal{N}}}(\mathbf{z}_{1},\dots, \mathbf{z}_{n})
&=&[\rho_{\varphi,\widehat{\theta }_{\mathcal{N}}}]^{n}\exp\left(\frac{1}{2}\sum_{i=1}^{n}\sum_{j=1}^{n}\mathcal{R}_{\mathbf{z}_{i}-\mathbf{z}_{j},\widehat{\theta }_{\mathcal{N}}}
(\varphi)(\varphi)\right),\nonumber\end{eqnarray}
\noindent for $\mathbf{z}_{1},\dots,\mathbf{z}_{n}\in \mathbb{Z}^{d},$ and  $\varphi\in \mathcal{H},$   with
\begin{eqnarray}
\rho_{\varphi ,\widehat{\theta }_{\mathcal{N}}} =\widehat{\rho_{\varphi,\theta }}=
\widehat{\rho^{(1)}_{\varphi ,\theta }}(\mathbf{z})&=&\exp\left(\frac{\mathcal{R}_{0,\widehat{\theta }_{\mathcal{N}}}(\varphi)(\varphi)}{2}\right),\quad \forall \mathbf{z}\in \mathbb{Z}^{d}.
\label{eqfoifes}
\end{eqnarray}
 \begin{theorem}
\label{th1ex} Let $\{\varkappa_{\mathbf{z}},\ \mathbf{z}\in \mathbb{Z}^{d}\}$ be a  spatial  stationary   zero--mean $\mathcal{H}$--valued  linear log--intensity random field. Let $\theta_{0}$ be the true parameter value lying  in  the interior  $\mbox{Int}(\Theta )$ of $\Theta .$     Under   \textbf{Assumptions A1--A2}, and conditions (\ref{cbn}) and
   (C1)--(C4),  the estimator (\ref{we2bb})--(\ref{wfs}) satisfies
\begin{eqnarray}
&&\widehat{\theta}_{\mathcal{N}}\to_{\mbox{a.s.}} \theta_{0},\quad \mathcal{\mathcal{N}}\to \infty,
 \label{convres}
\end{eqnarray}
\noindent where $\to_{\mbox{a.s.}}$ means the almost surely convergence.
\end{theorem}

\begin{proof} We consider here a compact version of the proof. Some additional technical details can be found in the Appendix. Specifically, the proof of this result follows from \textbf{Assumption A2} that implies,  for any orthonormal basis   $\left\{\varphi_{k},\ k\geq 1\right\}$   of $\mathcal{H}+i\mathcal{H},$ and  for $\mathbf{z}=(z_{1},\dots, z_{d})\in \mathbb{Z}^{d}$  with   $z_{i}\in [-\mathcal{N}_{i}+1,\mathcal{N}_{i}-1],$ $i=1,\dots,d,$\begin{eqnarray}
&&\sum_{k,l\geq 1}\left|\mathcal{C}(\mathbf{z},k,l)-\mathcal{R}_{\mathbf{z},\theta_{0}}(\varphi_{k})(\varphi_{l})\right|^{2}\to_{\mbox{a.s.}}0,\quad \mathcal{N}\to \infty,\label{fcshnormas}\end{eqnarray}
\noindent where the empirical Fourier coefficients $\mathcal{C}(\mathbf{z},k,l)$ of the periodogram operator are given by
\begin{eqnarray}
&&\mathcal{C}(\mathbf{z},k,l)=\int_{[-\pi,\pi]^{d}}\exp\left(i \sum_{j=1}^{d}\xi_{j}z_{j}
\right)\widetilde{X}^{\mathcal{N}}_{\boldsymbol{\xi }}(\varphi_{k})
\widetilde{X}^{\mathcal{N}}_{-\boldsymbol{\xi } }
(\varphi_{l})d\boldsymbol{\xi } \nonumber\\
&&=\frac{1}{\mathcal{N}}\left[\sum_{y_{1}=1}^{\mathcal{N}_{1}-z_{1}}\dots \sum_{y_{d}=1}^{\mathcal{N}_{d}-z_{d}} X_{\mathbf{y}}(\varphi_{k})X_{\mathbf{y}+\mathbf{z}}(\varphi_{l})\right],\quad |z_{i}|< \mathcal{N}_{i}, \quad i\in \{1,\dots,d\}\nonumber\\
&&\mathcal{C}(\mathbf{z},k,l)=0 ,\quad |z_{i}|\geq \mathcal{N}_{i}, \ \mbox{for some}\  i\in \{1,\dots,d\},\label{deffcro}
\end{eqnarray}
\noindent  for  $\mathbf{y}=(y_{1},\dots,y_{d})\in \prod_{i=1}^{d}[1,\mathcal{N}_{i}-z_{i}]\cap \mathbb{Z}^{d}.$
  Here, $\widetilde{X}^{\mathcal{N}}_{\boldsymbol{\xi }}(\varphi) =\langle \widetilde{\varkappa}^{\mathcal{N}}_{\boldsymbol{\xi }},\varphi\rangle,$  for every $\varphi\in \mathcal{H}+i\mathcal{H}.$

  Let  $M=\prod_{j=1}^{d}M_{j},$  under  (C4),  as $M\to \infty,$    the convergence of
\begin{eqnarray}&&q^{M}_{\boldsymbol{\xi },\theta}(\varphi_{k})(\varphi_{k})= \frac{1}{(2\pi)^{d}}\sum_{z_{j}\in [-M_{j}+1,M_{j}-1]; \ j=1,\dots,d}\exp\left(-i\sum_{j=1}^{d}\xi_{j}z_{j} \right)\nonumber\\ &&\hspace*{3cm}\times \prod_{j=1}^{d}\left(1-\frac{z_{j}}{M_{j}}\right)g(z_{1},\dots,z_{d},\theta,k)
 \label{eqccsis}
\end{eqnarray}

\noindent    to     $\mathcal{F}_{\boldsymbol{\xi },\theta}^{-1}(\varphi_{k})(\varphi_{k})$ holds  uniformly  in $k\geq 1,$ $\xi \in [-\pi,\pi]^{d},$ and $\theta \in \Theta .$ For each $\mathbf{z}= (z_{1},\dots,z_{d})\in \prod_{j=1}^{d} [-M_{j}+1,M_{j}-1],$ $g(\mathbf{z},\theta, k)=g(z_{1},\dots,z_{d},\theta, k)$ denotes the $\mathbf{z}$th--Fourier coefficient of $\mathcal{F}_{\boldsymbol{\varpi},\theta}^{-1}(\varphi_{k})(\varphi_{k})$  (see equation  (\ref{FCE}) in the Appendix).

 For $M$ sufficiently large, applying H\"older inequality,  \textbf{Assumption A2} and conditions (C1)--(C4)  lead to the a.s. convergence to zero,
 as $\mathcal{N}\to \infty,$  for any $\theta \in \Theta,$  of \begin{eqnarray}&&\hspace*{-0.7cm}\lim_{\mathcal{N}\to \infty}\sup_{k\geq 1}\int_{[-\pi,\pi]^{d}}
\left|q^{M}_{\boldsymbol{\xi },\theta}\left[\mathcal{I}_{\boldsymbol{\xi}}^{(\mathcal{N})}(\varphi_{k})(\varphi_{k})-\mathcal{F}_{\boldsymbol{\xi}
,\theta_{0} }(\varphi_{k})(\varphi_{k})\right]\right|d\boldsymbol{\xi}
\nonumber\\
  &&\hspace*{-0.5cm}\leq_{a.s.}\lim_{\mathcal{N}\to \infty} \sup_{k\geq 1}
\int_{[-\pi,\pi]^{d}}
\left|q^{M}_{\boldsymbol{\xi},\theta}\left[\mathcal{I}_{\boldsymbol{\xi}}^{(\mathcal{N})}(\varphi_{k})(\varphi_{k})-
\mathcal{F}_{\boldsymbol{\xi}
,\theta_{0} }(\varphi_{k})(\varphi_{k})\right]\right|^{2}d\boldsymbol{\xi}
 \nonumber \\
 &&\hspace*{-0.5cm}= \lim_{\mathcal{N}\to \infty} \sup_{k\geq 1}
 \frac{(2\pi)^{d}}{\mathcal{N}}\sum_{\mathbf{z}}
 \left|q^{M}_{\boldsymbol{\xi}_{\mathbf{z}},\theta}\mathcal{I}_{\xi_{\mathbf{z}}}^{(\mathcal{N})}(\varphi_{k})(\varphi_{k})-
 q^{M}_{\boldsymbol{\xi}_{\mathbf{z}},\theta}
 \mathcal{F}_{\boldsymbol{\xi}_{\mathbf{z}} ,\theta_{0} }(\varphi_{k})(\varphi_{k})\right|^{2}\underset{a.s}{=}0
 \label{eqlimbb}
 \end{eqnarray}

 \noindent (see \cite{Hannan73}, and the Appendix for more details).   Applying triangle inequality, under conditions (C3)--(C4),
   \begin{eqnarray}&&\sup_{k\geq 1}\int_{[-\pi,\pi]^{d}}\left|\mathcal{I}_{\boldsymbol{\xi}}^{(\mathcal{N})}\mathcal{F}_{\boldsymbol{\xi },\theta }^{-1}(\varphi_{k})(\varphi_{k})
- \mathcal{F}_{\boldsymbol{\xi},\theta_{0}}\mathcal{F}_{\boldsymbol{\xi},\theta }^{-1}(\varphi_{k})
(\varphi_{k})\right|
d\boldsymbol{\xi} \nonumber\\
&&\leq_{a.s.} \sup_{k\geq 1} \int_{[-\pi,\pi]^{d}}\left|\mathcal{I}_{\boldsymbol{\xi}}^{(\mathcal{N})}\left[\mathcal{F}_{\boldsymbol{\xi},\theta }^{-1}(\varphi_{k})(\varphi_{k})-q^{M}_{\boldsymbol{\xi},\theta }(\varphi_{k})(\varphi_{k})\right]\right|d\boldsymbol{\xi }\nonumber\\
&&+\sup_{k\geq 1} \int_{[-\pi ,\pi]^{d}}\left|q^{M}_{\boldsymbol{\xi},\theta }\left[\mathcal{I}_{\boldsymbol{\xi}}^{(\mathcal{N})}(\varphi_{k})(\varphi_{k})-\mathcal{F}_{\boldsymbol{\xi},\theta_{0}}
(\varphi_{k})(\varphi_{k})\right]\right|d\boldsymbol{\xi}\nonumber\\
&&+\sup_{k\geq 1}\int_{[-\pi,\pi]^{d}}\left|\mathcal{F}_{\boldsymbol{\xi},\theta_{0}}\left[q^{M}_{\boldsymbol{\xi},\theta }(\varphi_{k})(\varphi_{k})-\mathcal{F}_{\boldsymbol{\xi},\theta }^{-1}(\varphi_{k})(\varphi_{k})\right]\right|d\boldsymbol{\xi}\nonumber\\
&&\leq_{a.s} [2\pi]^{d} \varepsilon (M)\sup_{k\geq 1}\mathcal{C}(0,k,k)\nonumber\\
&&+\sup_{k\geq 1}\int_{[-\pi,\pi]^{d}}\left|q^{M}_{\boldsymbol{\xi},\theta }\left[ \mathcal{I}_{\boldsymbol{\xi}}^{(\mathcal{N})}(\varphi_{k})(\varphi_{k})-\mathcal{F}_{\boldsymbol{\xi},\theta_{0}}(\varphi_{k})(\varphi_{k})\right]\right|
d\boldsymbol{\xi}\nonumber\\
&&
+\varepsilon (M)\sup_{k\geq 1}\mathcal{R}_{0,\theta_{0}}(\varphi_{k})(\varphi_{k})\nonumber\\
&&= S_{1}(\mathcal{N},M)+S_{2}(\mathcal{N},M)+S_{3}(M),\quad \theta \in \Theta,
\label{ftthpe}\end{eqnarray}\noindent   with parameter $M$ denoting, as before,  the number of terms involved in the partial Cesaro sum of the Fourier series of  $\mathcal{F}_{\boldsymbol{\xi}, \theta }^{-1},$ and $\varepsilon (M)$ being the corresponding residual term.  Note that, under (C4),  the Fourier series of  $\mathcal{F}_{\boldsymbol{\xi}, \theta }^{-1}$ is Cesaro summable in $\mathcal{L}(\mathcal{H}+i\mathcal{H})$--norm,  uniformly in $(\boldsymbol{\xi},\theta )\in [-\pi,\pi]^{d}\times \Theta .$

 As $\mathcal{N}\to \infty,$ applying a.s. convergence to zero of equation (\ref{eqlimbb}), $$S_{2}(\mathcal{N},M)= \sup_{k\geq 1}\int_{[-\pi,\pi]^{d}}\left|q^{M}_{\boldsymbol{\xi},\theta }\left[ \mathcal{I}_{\boldsymbol{\xi}}^{(\mathcal{N})}(\varphi_{k})(\varphi_{k})-\mathcal{F}_{\boldsymbol{\xi},\theta_{0}}(\varphi_{k})(\varphi_{k})\right]\right|
d\boldsymbol{\xi} \to_{a.s.} 0.$$
 \noindent The Cesaro summability of the Fourier series of  $\mathcal{F}_{\boldsymbol{\xi}, \theta }^{-1}$ in  $\mathcal{L}(\mathcal{H}+i\mathcal{H})$--norm,  uniformly in $(\boldsymbol{\xi},\theta )\in [-\pi,\pi]^{d}\times \Theta $ in (C4) implies that, as $M\to \infty,$ $\varepsilon (M)\to 0.$  Finally, under \textbf{Assumption A2}, $\sup_{k\geq 1}\mathcal{C}(0,k,k)\to \sup_{k\geq 1}\mathcal{R}_{\mathbf{0},\theta_{0}}(\varphi_{k})(\varphi_{k}),$ as $\mathcal{N}\to \infty .$ Therefore, we obtain
\begin{eqnarray}
&&S_{1}(\mathcal{N},M)\to_{a.s.}[2\pi]^{d}\varepsilon (M)\sup_{k\geq 1}\mathcal{R}_{\mathbf{0},\theta_{0}}(\varphi_{k})(\varphi_{k}),\quad \mathcal{N}\to \infty\nonumber\\
&& \mbox{and}\quad \varepsilon (M)\sup_{k\geq 1}\mathcal{R}_{\mathbf{0},\theta_{0}}(\varphi_{k})(\varphi_{k})\to 0,\quad M\to \infty\nonumber\\
&&S_{2}(\mathcal{N},M)\to_{a.s.} 0\quad \mathcal{N}\to \infty\nonumber\\
&&S_{3}(M)= \varepsilon (M)\sup_{k\geq 1}\mathcal{R}_{\mathbf{0},\theta_{0}}(\varphi_{k})(\varphi_{k})\to 0,\quad
\quad  M\to \infty.
\label{cvzer}
\end{eqnarray}
Hence, from  (\ref{ftthpe})--(\ref{cvzer}),  we obtain, for $\theta \in \Theta,$
\begin{eqnarray}&&\sigma_{\mathcal{N}}(\theta )\to_{\mbox{a.s.}}\sup_{k\geq 1}\frac{1}{(2\pi)^{d}}\int_{[-\pi,\pi]^{d}}\mathcal{F}_{\boldsymbol{\xi},\theta_{0}}\mathcal{F}_{\boldsymbol{\xi },\theta }^{-1}(\varphi_{k}) (\varphi_{k})
d\boldsymbol{\xi}\nonumber\\ &&= \sup_{k\geq 1} \frac{1}{(2\pi)^{d}}\int_{[-\pi,\pi]^{d}}\mathcal{F}_{\boldsymbol{\xi },\theta }^{-1}\mathcal{F}_{\boldsymbol{\xi},\theta_{0}}(\varphi_{k}) (\varphi_{k})
d\boldsymbol{\xi},\quad \mathcal{N}\to \infty. \label{limas}
\end{eqnarray}

Under (C3), \begin{equation} \sup_{k\geq 1}\frac{1}{(2\pi)^{d}}\int_{[-\pi,\pi]^{d}}\mathcal{F}_{\boldsymbol{\xi},\theta_{0}}\mathcal{F}_{\boldsymbol{\xi} ,\theta}^{-1}(\varphi_{k})(\varphi_{k})d\boldsymbol{\xi}>1,\quad \theta\neq \theta_{0},\label{eqineqhh}\end{equation}
\noindent since, for each $k\geq 1,$   $\mathcal{F}_{\boldsymbol{\xi} ,\theta_{0}}\mathcal{F}_{\boldsymbol{\xi} ,\theta}^{-1}(\varphi_{k})(\varphi_{k})$ defines the spectrum of a spatial stationary linear process with unit one--step prediction variance (see Remarks \ref{rem4pf}--\ref{uospv}).
 Specifically, the variance of such a   process is given by\linebreak
  $\int_{[-\pi,\pi]^{d}}\mathcal{F}_{\boldsymbol{\xi} ,\theta_{0}}\mathcal{F}_{\boldsymbol{\xi} ,\theta}^{-1}(\varphi_{k})(\varphi_{k})d\boldsymbol{\xi} ,$
  which must be  larger than  the one--step prediction variance,  unless the spectrum is constant for every $k\geq 1.$ This fact does not hold because it means  that the composition
$\mathcal{F}_{\boldsymbol{\xi} ,\theta_{0}}\mathcal{F}_{\boldsymbol{\xi} ,\theta}^{-1}$  of operators   $\mathcal{F}_{\boldsymbol{\xi},\theta_{0}}$ and $\mathcal{F}_{\boldsymbol{\xi} ,\theta}^{-1}$ coincides with the identity operator on $\mathcal{H}+i\mathcal{H},$ which is not possible in view of identifiability condition  (C1), since $\theta \neq \theta_{0}$ in (\ref{eqineqhh}). Thus,
\begin{equation}
 \inf_{\theta\in \Theta}\sup_{k\geq 1}\frac{1}{(2\pi)^{d}}\int_{[-\pi,\pi]^{d}}\mathcal{F}_{\boldsymbol{\xi } ,\theta_{0}}\mathcal{F}_{\boldsymbol{\xi } ,\theta}^{-1}(\varphi_{k})(\varphi_{k})d\boldsymbol{\xi } =1.\label{infbb}\end{equation}
 \noindent From  (\ref{infbb}), the theoretical counterpart of the empirical  loss function  (\ref{wfs})   attaches the minimum at $\theta =\theta_{0}.$ The strong consistency of the derived parameter estimator follows from this fact, and from equations (\ref{limas}) and (\ref{eqineqhh}) (see Appendix for more details).

\end{proof}
\section{Simulation study}
\label{simulationstdy}  This section illustrates the derived strong consistency  of the parameter  estimator (\ref{we2bb})--(\ref{wfs}), as well as its  mean--square consistency. We restrict our attention  to the case where the log--intensity random field is a   SAR$\mathcal{H}$(1) process (see \cite{Ruiz11a}). As commented in the Introduction,   in the numerical examples below we address the problem of estimating the hyperparameters, characterizing the point spectra and eigenvectors of the spatial correlation operators, that are also involved in the parameterization of our  spectral density operator family (see  equation (\ref{eqsdosarh1}) below).

Let, as before,  $\mathcal{H}=L^{2}(\mathcal{T})$ be the space of square--integrable functions on the interval $\mathcal{T}\subseteq \mathbb{R}_{+}.$ Consider the case $d=2,$ and  assume that the $\mathcal{H}$--valued  log--intensity  random field $\{\varkappa_{\mathbf{z}},\ \mathbf{z}\in \mathbb{Z}^{2}\}$ satisfies the  SAR$\mathcal{H}$(1) state equation

\begin{equation}
\varkappa_{i,j}= L_{1}(\varkappa_{i-1,j})+L_{2}(\varkappa_{i,j-1})+L_{3}(\varkappa_{i-1,j-1})+\epsilon_{i,j},\quad (i,j)\in \mathbb{Z}^{2},\label{Xij}
\end{equation}
\noindent where $\epsilon=\{\epsilon_{i,j}, (i,j)\in \mathbb{Z}^2\}$ denotes  a spatial  $\mathcal{H}$--valued  zero--mean  generalized white noise. Hence, for any spatial node  $(i,j),$   $E|\epsilon_{i,j}(\varphi_{k})|^{2}=\left\|\varphi_{k}\right\|^{2}=\sigma_{k}^{2}=1,$ for  $k\geq 1,$  with $\{\varphi_{k},\ k\geq 1\}$ being  an orthonormal basis  of $\mathcal{H}.$   Thus, $\mathcal{R}_{i,j}^{\epsilon }=E\left(\epsilon_{i+k,j+l}\otimes \epsilon_{k,l}\right)=E\left(\epsilon_{i,j}\otimes \epsilon_{0,0}\right)=I_{\mathcal{H}},$   for every   $(i,j), (k,l) \in \mathbb{Z}^{2},$ with $I_{\mathcal{H}}$ denoting the identity operator on $\mathcal{H}.$
 We  work under the assumption  of  $\{\epsilon_{i,j},\ (i,j)\in \mathbb{Z}^{2}\}$ being   Gaussian. Specifically,  $\{\epsilon_{i,j}(\varphi_{k}),\ k\geq 1,\ (i,j)\in \mathbb{Z}^{2}\}$ is a family of  independent standard  Gaussian random variables. For every $(i,j)\in \mathbb{Z}^{2},$  $\epsilon_{i,j}$ is uncorrelated with $\varkappa_{i-1,j},$ $\varkappa_{i,j-1},$  $\varkappa_{i-1,j-1}.$ The pure point spectra  of $L_{i},$ $i=1,2,3,$ are defined from the  pure point spectra  of the linear operators involved in the spatial functional infinite--order  moving average  (SMA$\mathcal{H}$($\infty$)) representation of  $\varkappa_{i,j}.$

 In the subsequent examples, $L_{i},$ $i=1,2,3,$   will be introduced in terms of the eigenvectors $\{\phi_{k},\ k\geq 1\}$  of  the Dirichlet negative Laplacian operator $(-\Delta )_{\mathcal{T}}$ on an interval $\mathcal{T}$
    \begin{eqnarray}
&&\phi_{k}(t)=\sin\left(\frac{\pi t k}{T}\right),\quad t\in \mathcal{T}=[0,T],\quad k\geq 1.\label{eigenvectors}
\end{eqnarray}
   Applying Spectral Theorem on Spectral Calculus for self--adjoint operators on a separable Hilbert space (see, e.g.,  \cite{Dutray1985}, pp. 112--126), we can characterize the pure point spectrum of any  continuous function
 of  $(-\Delta )_{\mathcal{T}},$ by applying this function to the eigenvalues of  $(-\Delta )_{\mathcal{T}}.$
 This is the framework adopted in Examples 1 and 2 below for the introduction of operators $L_{i},$ $i=1,2,3.$
On the other hand, in the  case of considering  that the functional values of the  $\mathcal{H}$--valued log--intensity process are supported on the entire real line, these models can be extended by considering  the weak--sense Fourier transform of Riesz and Bessel potentials (see, e.g., \cite{Triebel78}).

\medskip

\noindent\emph{Example 1}.  In this example,  we address  the problem of estimating parameter $\theta \in \Theta $ characterizing  the support $\mathcal{T}(\theta )=[0,\theta ]$ of the eigenvectors (\ref{eigenvectors}). We thus have   $L_{i,\theta},$ $i=1,2,3,$  given by

\begin{eqnarray}
  &&L_{1,\theta }=  \Phi_{M_{1}(\theta )}\mathcal{K}_{1}(\theta ,\pi)(-\Delta)_{\mathcal{T}(\theta )}^{-\beta_{1} },
 \quad L_{2,\theta }=\Phi_{M_{2}(\theta )}\mathcal{K}_{2}(\theta ,\pi)(-\Delta)_{\mathcal{T}(\theta )}^{-\beta_{2} },\nonumber\\
  && L_{3,\theta }=  -L_{1,\theta}L_{2,\theta}= -L_{2,\theta}L_{1,\theta},\label{autocorrsarh1}
  \end{eqnarray}

\noindent where $\Phi_{M_{i}(\theta )}$  denotes the projection operator onto the subspace generated by the eigenvectors $\phi_{1},\dots, \phi_{M_{i}(\theta )}$  of $(-\Delta)_{\mathcal{T}(\theta )},$  with $M_{i}(\theta )\geq 1$  being a bounded function in $\theta \in \Theta ,$  for $i=1,2.$  Here,  $\beta_{i}, \mathcal{K}_{i}\in \mathbb{R}_{+}$ are known, and   $\mathcal{K}_{i}$ is continuous in $\theta \in \Theta,$ for $i=1,2.$
Applying Spectral Theorem on Spectral Calculus for self--adjoint operators on a separable Hilbert space (see, e.g.,  \cite{Dutray1985}),
the eigenvalues $\{\lambda_{k,i}(\theta ), k\geq 1\}$  of the autocorrelation operator $L_{i,\theta }$ are given,  for   $i=1,2,3,$ by
 \begin{eqnarray}
&&\lambda_{k,1}(\theta )=1_{[1,M_{1}(\theta )]}(k)\mathcal{K}_{1}(\theta ,\pi)\left[\lambda_{k}\left((-\Delta)_{\mathcal{T}(\theta )}\right)\right]^{-\beta_{1}}\nonumber\\
&& \lambda_{k,2}(\theta )=1_{[1,M_{2}(\theta )]}(k)\mathcal{K}_{2}(\theta , \pi)\left[\lambda_{k}\left((-\Delta)_{\mathcal{T}(\theta )}\right)\right]^{-\beta_{2}}
\nonumber\\
&&\lambda_{k,3}(\theta )= -\lambda_{k,1}(\theta ) \lambda_{k,2}(\theta ),\  k\geq 1,\ \theta \in \Theta,\label{mss}
\end{eqnarray}
 \noindent where   $1_{[1,M_{i}(\theta )]}(k)$    denotes the indicator function of the set  $[1,M_{i}(\theta )]\subset \mathbb{N},$ with, as before,  $M_{i}(\theta )\geq 1,$    for  $i=1,2.$  For $k\geq 1,$ $\lambda_{k}\left((-\Delta)_{\mathcal{T}(\theta )}\right)$ denotes the $k$--th eigenevalue of $(-\Delta)_{\mathcal{T}(\theta )}.$ Consider the particular case $M_{1}(\theta)=M_{2}(\theta)=M(\theta),$
      and assume  that $\lambda_{k,i}(\theta ),$ $k\geq 1,$ $i=1,2,3,$ are such that  the roots of the polynomial $P_{k,\theta }(z_{1},z_{2})=\sum_{i=0}^{1}\sum_{j=0}^{1}p_{k,i,j,\theta }z_{1}^{i}z_{2}^{j},$ are outside  the unit polydisc (i.e.,  the product of  two open unit disks in the complex plane). Here,  $p_{k,0,0,\theta }=1,$ $p_{k,1,0,\theta }=\lambda_{k,1}(\theta ),$ $p_{k,0,1,\theta }=\lambda_{k,2}(\theta ),$ and $p_{k,1,1,\theta }=  -\lambda_{k,1}(\theta ) \lambda_{k,2}(\theta ),$ $k\geq 1.$ From equations (\ref{Xij})--(\ref{mss}),  we compute the  spectral density operator, given by, for $\boldsymbol{\xi}\in [-\pi,\pi]^{2},$
    \begin{eqnarray}&&\mathcal{F}_{\boldsymbol{\xi},\theta }(\phi_{k,\theta})(\phi_{k,\theta})=\nonumber\\
        &&\hspace*{-0.5cm}= \frac{1}{\left|1-\lambda_{k,1}(\theta )\exp(i\xi_{1})-\lambda_{k,2}(\theta )\exp(i\xi_{2})-\lambda_{k,3}(\theta )\exp(i(\xi_{1}+\xi_{2}))\right|^{2}}\label{eqsdosarh1}\\
 &&L_{q,\theta }(g)(h)=\sum_{k=1}^{\infty}\lambda_{k,q}(\theta )\left\langle \phi_{k, \theta},h\right\rangle\left\langle \phi_{k, \theta},
g\right\rangle,\quad \forall g,h\in \mathcal{H},\ q=1,2,3\nonumber\\
&&\phi_{k,\theta }(t)=\sin\left(\frac{\pi t k}{\theta }\right),\quad t\in \mathcal{T}(\theta )=[0,\theta ],\ k\in [1, M(\theta )],\  \theta \in \Theta .\nonumber
\end{eqnarray}

\begin{figure}[hptb]
\begin{center}
\includegraphics[width=4.5cm,height=4.5cm]{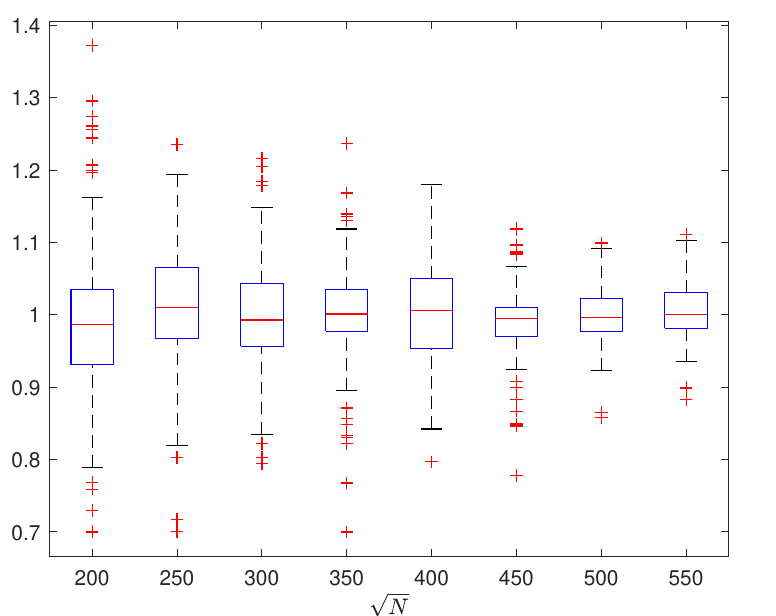}
\includegraphics[width=4.5cm,height=4.5cm]{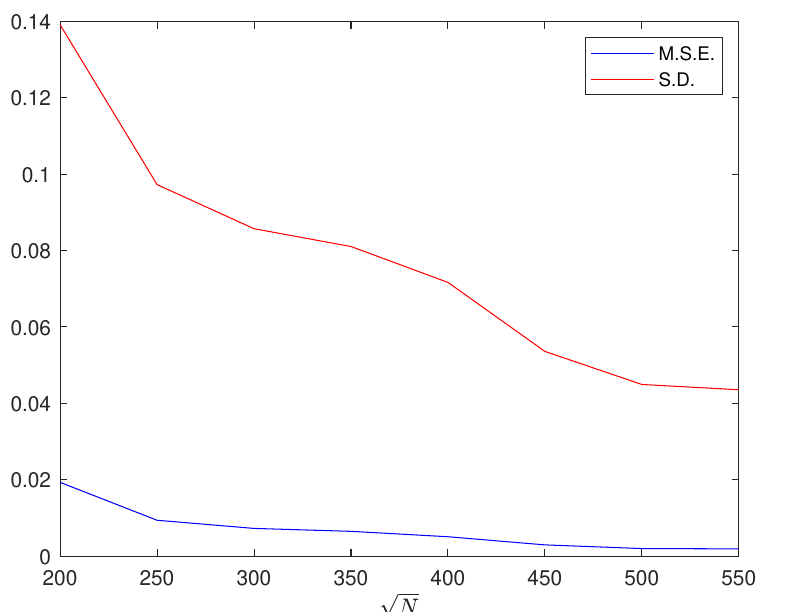}
\end{center}
\caption{Boxplots of the  sample values of   $\widehat{\theta}_{\mathcal{N}},$  for $\theta_{0}= 1$ (left--plot).
Empirical Mean Square Errors (M.S.E), blue line,  and Standard Deviations (S.D.), red line.}
\label{figs1}
\end{figure}

Under the  conditions  assumed   on model parameters (\ref{mss}),   \textbf{Assumption A2} is satisfied in our SAR$\mathcal{H}$(1)  framework,  as follows from Corollary 2.3 in \cite{Bosq2000}, applying similar results to those obtained in  Corollary 4.1 and  Theorem 4.8 of this monograph.
 Parameter $\theta  $  univocally defines  the support $\mathcal{T}(\theta )$ of the eigenvectors of the Dirichlet
    negative Laplacian operator. Hence,  the identifiability condition (C1) is satisfied.    Conditions  (\ref{cbn}), (C2) and (C4) are  obtained from equations (\ref{mss})--(\ref{eqsdosarh1}), under
the restrictions considered  on the parametric coefficients  $\{\lambda_{k,i}(\theta ),\ k\in [1, M(\theta )],\ \theta \in \Theta,\ i=1,2\},$  defining the  polynomial family   $\left\{P_{k,\theta },\ k  \in [1,M(\theta )],\ \theta \in \Theta \right\}.$
  Condition (C3)  also holds since  $\sigma_{k}^{2}=1,$ for every $k\geq 1.$
    The  strong--consistency  of the SAR$\mathcal{H}$(1) plug--in parametric predictor
\begin{equation}\widehat{\varkappa}_{i,j}^{\mathcal{N}}(\theta )=L_{1,\widehat{\theta }_{\mathcal{N}}}\varkappa_{i-1,j}+L_{2,\widehat{\theta }_{\mathcal{N}}}\varkappa_{i,j-1}+L_{3,\widehat{\theta }_{\mathcal{N}}}\varkappa_{i-1,j-1}, \quad \forall (i,j)\in \mathbb{Z}^{2},
\label{SARHpp}
\end{equation}
\noindent  is then obtained under similar conditions to those  assumed in \cite{Bosq2000} in the AR$\mathcal{H}$(1) framework.
The  estimation results  are displayed in Table \ref{t1} and Figure \ref{figs1}, where we have considered $\theta_{0} =1,$
$\mathcal{K}_{1}(\theta ,\pi)=\frac{\theta^{2-11/10}}{\pi^{2-11/10}},$ $\mathcal{K}_{2}(\theta , \pi)=\frac{\theta^{2-12/10}}{\pi^{2-12/10}},$ $\beta_{1}=\frac{11}{20},$ $\beta_{2}=\frac{12}{20},$ $M(\theta )=[10\theta  ],$ and  $\theta \in \Theta =[0.7, 4],$ with $[\cdot]$ denoting the integer part function.
We have tested the   spatial functional sample sizes $\mathcal{N}=40000,       62500,       90000,      122500, 160000,      202500,      250000,$     $302500.$
Specifically, Figure \ref{figs1}--left displays the sample values of  $\widehat{\theta}_{\mathcal{N}},$ based on $100$ generations of each functional sample. The empirical mean quadratic errors are also computed
in Table \ref{t1} and Figure \ref{figs1}--right.

\begin{table}
\caption{Location parameter estimates. Sample Mean, Standard Deviation, and Empirical Mean Square Errors (M.S.E.), from $100$ generations,  $\theta_{0}= 1$ }
\label{t1}
\begin{center}
\begin{tabular}{cccc}
\hline
$\mathcal{N}$&$\overline{\widehat{\theta}_{\mathcal{N}}}$&$\sigma(\widehat{\theta}_{\mathcal{N}})$&M.S.E\\\hline
40000&0.9867 & 0.1391  &0.0193\\
62500&1.0066 & 0.0972  &0.0094\\
90000&0.9993 & 0.0857  &0.0073\\
122500&1.0008 & 0.0810  &0.0065\\
160000&1.0034 & 0.0716  &0.0051\\
202500&0.9893 & 0.0536  &0.0030\\
250000&0.9995 & 0.0449  &0.0020\\
302500&1.0060 & 0.0436  &0.0019\\
\hline
\end{tabular}
\end{center}
\end{table}

\begin{figure}[hptb]
\begin{center}
\includegraphics[width=4.5cm,height=4.5cm]{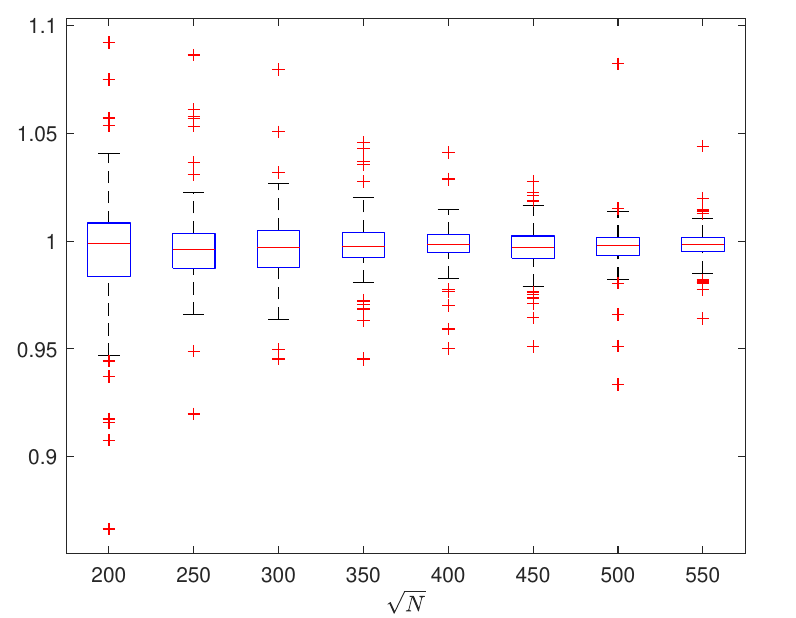}
\includegraphics[width=4.5cm,height=4.5cm]{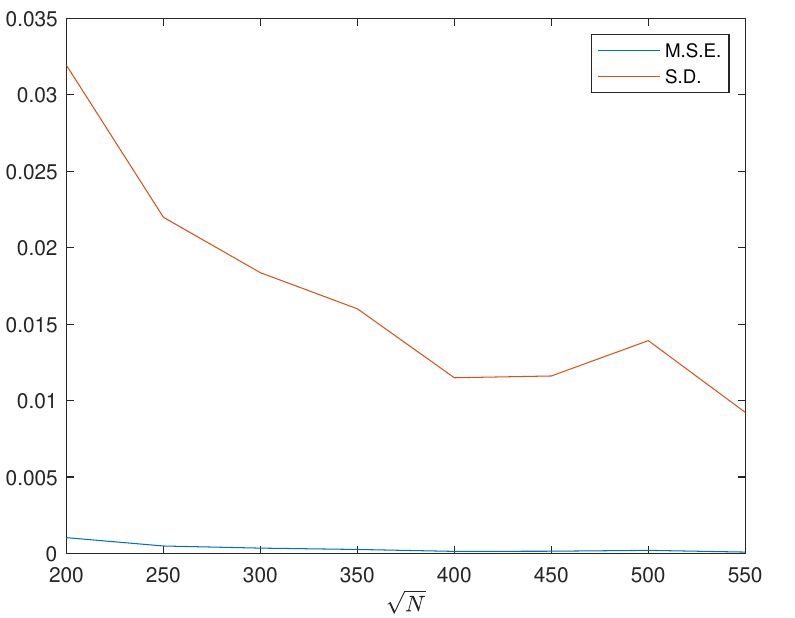}
\end{center}
\caption{Boxplots of the sample values of  $\widehat{\theta}_{1,1,\mathcal{N}},$ for $\theta_{1,1,0}= 1,$ based on $100$ generations  (left--plot).
Empirical Mean Square Errors (M.S.E.),   blue line, and  Sample Standard Deviation (S.D.), red line  (right--plot)}
\label{fig2s}
\end{figure}

\begin{figure}[hptb]
\begin{center}
\includegraphics[width=4.5cm,height=4.5cm]{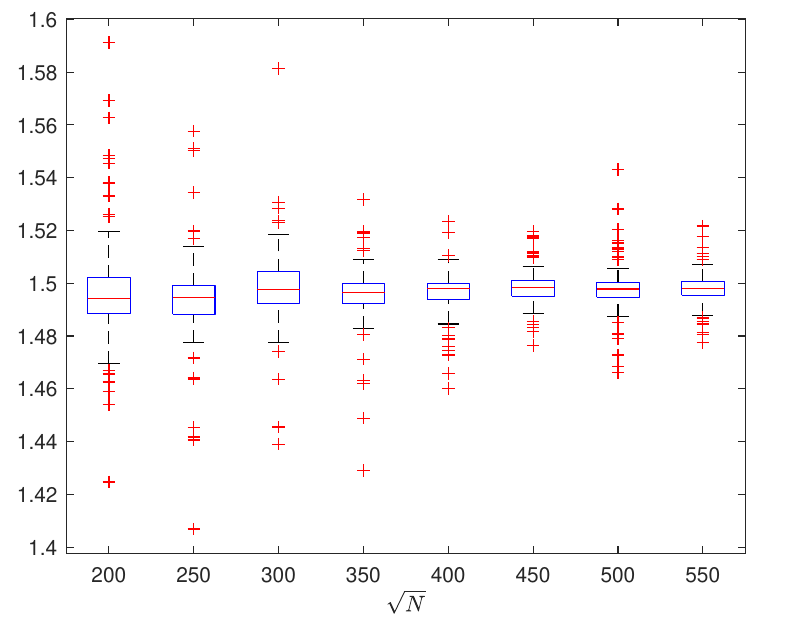}
\includegraphics[width=4.5cm,height=4.5cm]{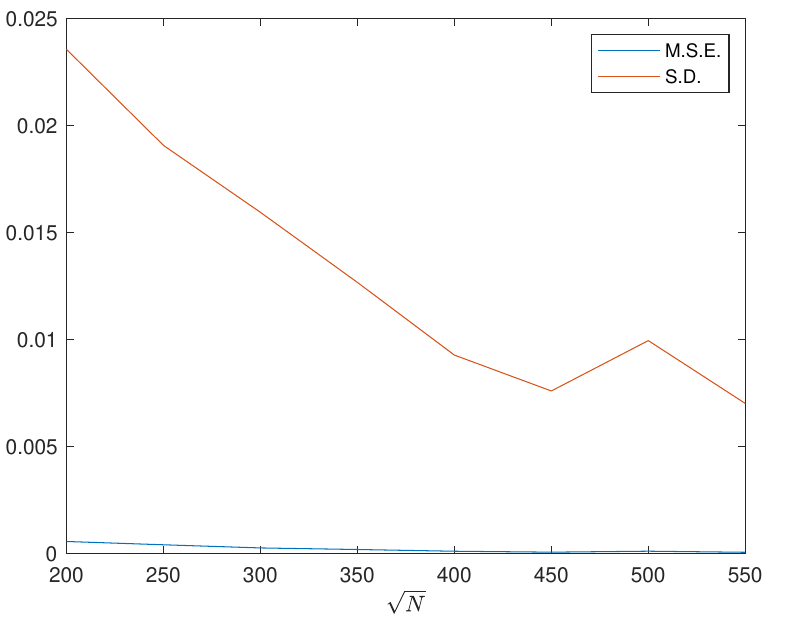}
\end{center}
\caption{Boxplots of the sample values of   $\widehat{\theta}_{2,1,\mathcal{N}},$ for $\theta_{2,1,0}= 1.5,$ based on $100$ generations  (left--plot).
Empirical Mean Square Errors (M.S.E.),   blue line, and Sample  Standard Deviation (S.D.),  red line (right--plot)}
\label{fig3s}
\end{figure}

\begin{figure}[hptb]
\begin{center}
\includegraphics[width=4.5cm,height=4.5cm]{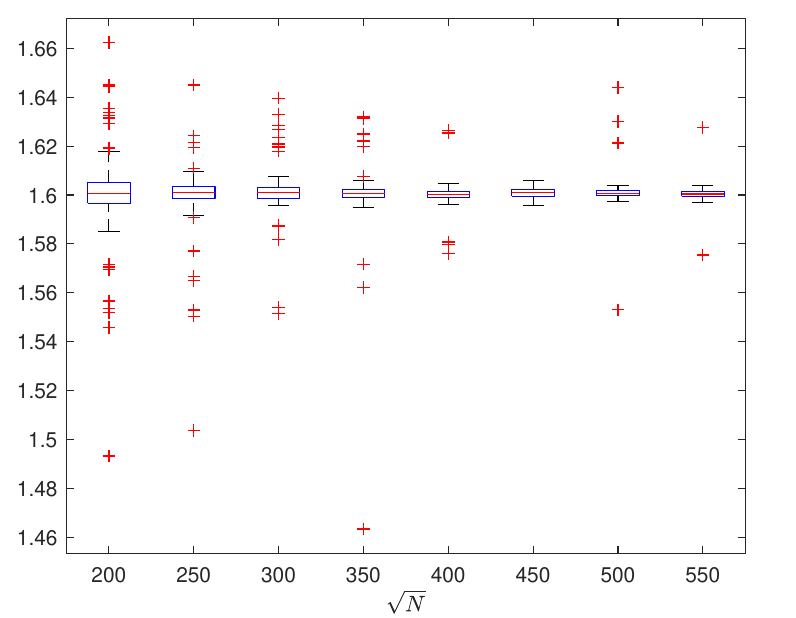}
\includegraphics[width=4.5cm,height=4.5cm]{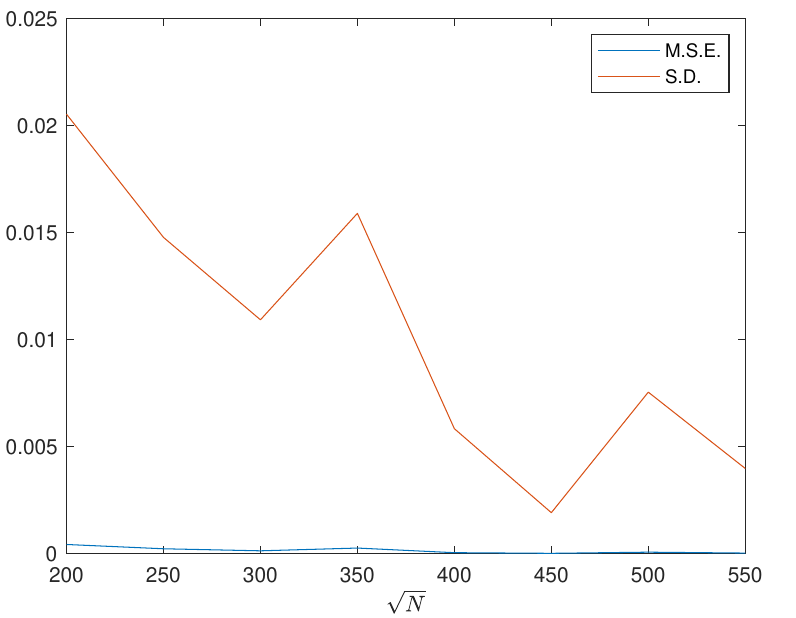}
\end{center}
\caption{Boxplots of sample values of  $\widehat{\theta}_{1,2,\mathcal{N}},$ for $\theta_{1,2,0}= 1.6$,  based on $100$ generations (left--plot).
Empirical Mean Square Errors (M.S.E.),   blue line, and Sample  Standard Deviation (S.D.),  red line (right--plot)}
\label{fig4s}
\end{figure}

\begin{figure}[hptb]
\begin{center}
\includegraphics[width=4.5cm,height=4.5cm]{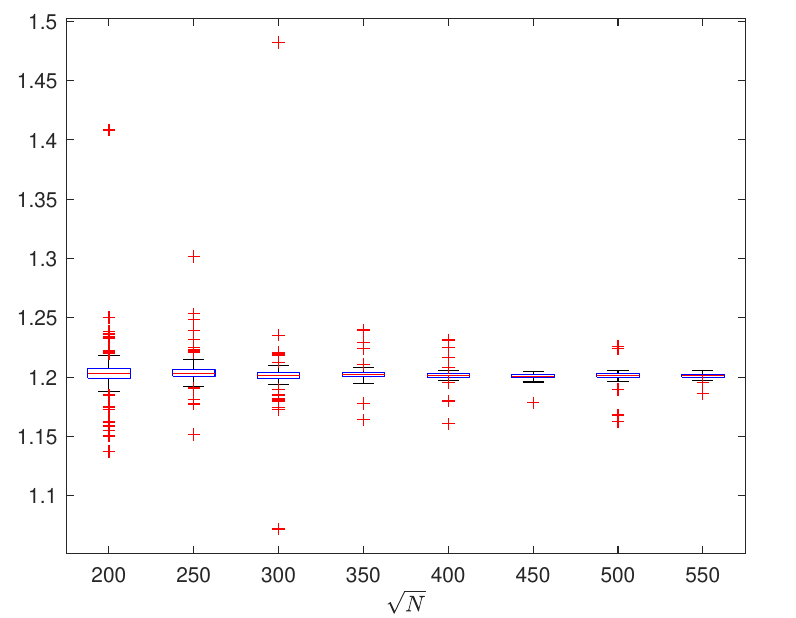}
\includegraphics[width=4.5cm,height=4.5cm]{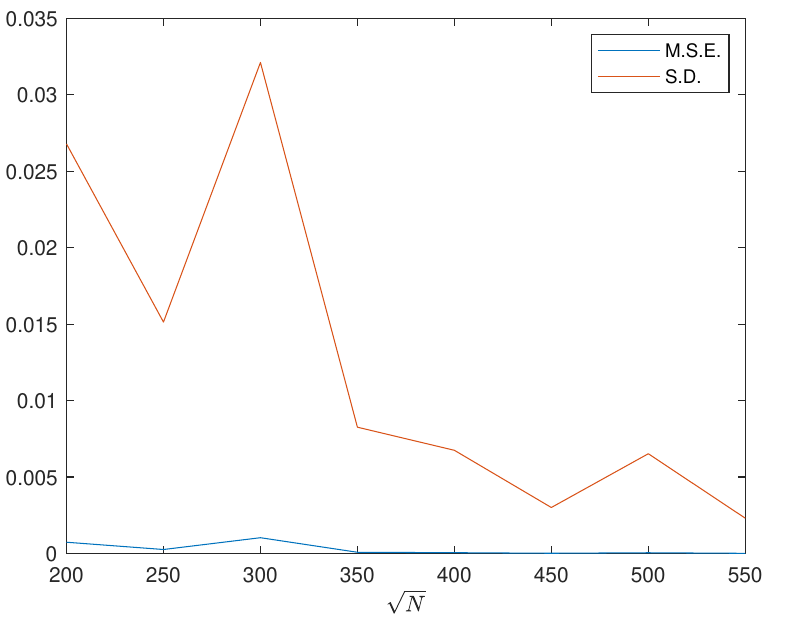}
\end{center}
\caption{Boxplots of sample values of  $\widehat{\theta}_{2,2,\mathcal{N}},$ for  $\theta_{2,2,0}= 1.2,$  based on $100$ generations  (left--plot).
Empirical Mean Square Errors (M.S.E.),   blue line, and Sample  Standard Deviation (S.D.),  red line (right--plot)}
\label{fig5s}
\end{figure}

\noindent \emph{Example 2}.  Let now consider the following parametric family of autocorrelation operators
   $L_{q,\boldsymbol{\theta}_{q}},$  $q=1,2,$
\begin{eqnarray}
  &&L_{1,\boldsymbol{\theta}_{1} }=\Phi_{M}\theta_{1,1}\left[ \theta_{1,2}+ (1/\pi)(-\Delta )^{1/2}_{[0,1]}\right]^{-1}\nonumber\\
  &&\quad L_{2,\boldsymbol{\theta}_{2} }=\Phi_{M}
 \theta_{2,1}\left[ \theta_{2,2}+ (1/\pi)(-\Delta )^{1/2}_{[0,1]}\right]^{-1},\quad (\boldsymbol{\theta} _{1},  \boldsymbol{\theta}_{2})\in \Theta \nonumber\\
 &&  L_{3,\boldsymbol{\theta}_{3}  }=  -L_{1,\boldsymbol{\theta}_{1} }L_{2, \boldsymbol{\theta}_{2} }= -L_{2, \boldsymbol{\theta}_{2} }L_{1,\boldsymbol{\theta}_{1}}\nonumber\\
&&\phi_{k}(t)=\sin\left(\pi t k\right),\quad t\in [0,1],\quad k\geq 1  \nonumber\\
&&\lambda_{k,q}(\boldsymbol{\theta }_{q})= 1_{[1,M]}(k)\frac{\theta_{q,1}}{k + \theta_{q,2}},\ q=1,2,\  \lambda_{k,3}(\boldsymbol{\theta }_{3})=-\lambda_{k,1}(\boldsymbol{\theta }_{1})\lambda_{k,2}(\boldsymbol{\theta }_{2}), \quad k\geq 1\nonumber\\
  \label{mss2}
\end{eqnarray}
\noindent where, as before, $1_{[1,M]}(k)$ is the indicator function of the set $[1,M]\subset \mathbb{N},$  $\Phi_{M},$ $M\geq 1,$ denotes the projection operator onto the subspace generated by the eigenvectors $\phi_{1},\dots,\phi_{M}$ of Dirichlet negative  Laplacian operator $(-\Delta )_{[0,1]}$ on the interval $[0,1].$     Here,  $\{\lambda_{k,1}(\boldsymbol{\theta }_{1}),\ \lambda_{k,2}(\boldsymbol{\theta }_{2}),\  k\geq 1,\ (\boldsymbol{\theta }_{1},\boldsymbol{\theta}_{2})\in \Theta\}$  is the parametric family of   eigenvalues of the autocorrelation operators  \linebreak $\{L_{1,\boldsymbol{\theta}_{1} },\ L_{2,\boldsymbol{\theta}_{2} },\  (\boldsymbol{\theta }_{1},\boldsymbol{\theta}_{2})\in \Theta\}.$
Under similar conditions to Example 1 on these parametric families of eigenvalues of   $\{L_{1,\boldsymbol{\theta}_{1} },\ L_{2,\boldsymbol{\theta}_{2} },\  (\boldsymbol{\theta }_{1},\boldsymbol{\theta}_{2})\in \Theta\},$ Assumptions A1--A2, and conditions (\ref{cbn}) and (C1)--(C4) hold.
Thus, the strong--consistency of the parameter estimator $\widehat{\theta }_{\mathcal{N}}$ follows from
Theorem  \ref{th1ex}. Particularly,   we have considered  the   scale parameter values $\theta_{1,1} = 1$, $\theta_{2,1} = 1.5$, and  the location parameters values $\theta_{1,2}=1.6$, and  $\theta_{2,2}=1.2,$ lying in the interior of $\Theta =([0.7,1.3]\times [1.3, 1.9])\times ([1.2, 1.8]\times  [0.9, 1.5]).$
 See the numerical results  displayed at the left--plots of Figures \ref{fig2s}--\ref{fig5s}, from  $100$ generations of each one of the spatial functional samples of sizes  $\mathcal{N}=40000,       62500,       90000,      122500, 160000,$      $202500,      250000,      302500.$ The empirical mean quadratic errors have  also been computed for these  sample sizes  (see right--plots of  Figures \ref{fig2s}--\ref{fig5s} and
 Table \ref{t2}).
It can also be observed a good performance of the proposed estimation methodology, for the considered spatial functional samples sizes, and the chosen truncation parameter value  $M=10.$  Note that this truncation parameter value provides the  threshold  dimension to ensure a stable behavior of the parameter vector estimator, with respect to the  input spatial functional data, according to the temporal discretization step size and  smoothing considered.
\begin{table}
\caption{Location and scale parameter estimates. Sample Mean, Standard Deviation and Empirical Mean Quadratic Errors (M.S.E.), from $100$ generations, $\theta_{1,1,0} = 1$, $\theta_{2,1,0} = 1.5$, $\theta_{1,2,0}=1.6$, and  $\theta_{2,2,0}=1.2.$}
\label{t2}
\begin{center}
\begin{tabular}{cccc}
\hline
$\mathcal{N}$&$\overline{\widehat{\theta}_{1,1,\mathcal{N}}}$&$\sigma(\widehat{\theta}_{1,1,\mathcal{N}})$&M.S.E.\\\hline
40000&0.9955 & 0.0319  &0.00103\\
62500&0.9975 & 0.0220  &0.00048\\
90000&0.9965 & 0.0184  &0.00035\\
122500&0.9978 & 0.0160  &0.00026\\
160000&0.9984 & 0.0115  &0.00013\\
202500&0.9969 & 0.0116  &0.00014\\
250000&0.9973 & 0.0139  &0.00020\\
302500&0.9983 & 0.0092  &0.00009\\
\hline
\end{tabular}
\end{center}
\begin{center}
\begin{tabular}{cccc}
\hline
$\mathcal{N}$ & $\overline{\widehat{\theta}_{2,1,\mathcal{N}}}$ & $\sigma(\widehat{\theta}_{2,1,\mathcal{N}})$ & M.S.E.\\\hline
40000&1.4972 & 0.0236  &0.00056\\
62500&1.4934 & 0.0191  &0.00040\\
90000&1.4980 & 0.0159  &0.00026\\
122500&1.4951 & 0.0127  &0.00018\\
160000&1.4959 & 0.0093  &0.00010\\
202500&1.4989 & 0.0076  &0.00006\\
250000&1.4977 & 0.0099  &0.00010\\
302500&1.4979 & 0.0070  &0.00005\\
\hline
\end{tabular}
\end{center}
\begin{center}
\begin{tabular}{cccc}
\hline
$\mathcal{N}$&$\overline{\widehat{\theta}_{1,2, \mathcal{N}}}$&$\sigma(\widehat{\theta}_{1,2,\mathcal{N}})$&M.S.E.\\\hline
40000&1.6003 & 0.0205  &0.00042\\
62500&1.5991 & 0.0148  &0.00022\\
90000&1.6018 & 0.0109  &0.00012\\
122500&1.5998 & 0.0159  &0.00025\\
160000&1.6001 & 0.0058  &0.00003\\
202500&1.6008 & 0.0019  &0.00000\\
250000&1.6011 & 0.0075  &0.00006\\
302500&1.6005 & 0.0040  &0.00002\\
\hline
\end{tabular}
\end{center}
\begin{center}
\begin{tabular}{cccc}
\hline
$\mathcal{N}$&$\overline{\widehat{\theta}_{2,2,\mathcal{N}}}$&$\sigma(\widehat{\theta}_{2,2,\mathcal{N}})$&M.S.E.\\\hline
40000&1.2046 & 0.0268  &0.00073\\
62500&1.2054 & 0.0151  &0.00026\\
90000&1.2024 & 0.0321  &0.00103\\
122500&1.2027 & 0.0083  &0.00007\\
160000&1.2015 & 0.0067  &0.00005\\
202500&1.2005 & 0.0030  &0.00001\\
250000&1.2008 & 0.0065  &0.00004\\
302500&1.2008 & 0.0023  &0.00001\\
\hline
\end{tabular}
\end{center}
\end{table}


 \section{Real-data example}
\label{s6} Data are reported by  the Spanish National Statistical Institute, consisting of 432 monthly records  on  the number of respiratory disease mortality cases at the  48  Spanish  provinces in the Iberian Peninsula, during the period 1980--2015.
  A spatial heterogeneous 10th--degree polynomial trend model fitting is achieved  after  suitable preprocessing of our original data set. Spatial functional residual  correlation analysis is performed in the spectral domain, in terms of the SAR$\mathcal{H}$(1) model introduced in the previous section.  Specifically, the main steps of the  implemented estimation algorithm   are the following.

\begin{itemize}
\item[Step 1.]    The  cumulative number of mortality cases at  the $48$  Spanish  provinces, in the period 1980--2015, are computed from  the 432 monthly records.
\item[Step 2.]  The step cumulative mortality  curves in Step 1 are interpolated   at $1725$  temporal   nodes,  and cubic B--spline smoothed.
\item[Step 3.]  The curve data in Step 2 are spatially interpolated to a $\mathcal{N}=20\times 20$  regular  grid,  by applying  Inverse Distance Weighting interpolation.
\item[Step 4.] The derivatives and  logarithm transforms of the curve data in Step 3 provide the functional   values of the log--intensity at the spatial nodes of the  $\mathcal{N}= 20\times 20$   regular  grid.
\item[Step 5.]     Least-squares polynomial fitting is implemented to approximate the log--intensity  curve trend at each spatial node.
\item[Step 6.]   The residual log--intensity curve  values obtained in Step 5 are projected onto the orthonormal basis $\{\sin\left(\pi p t/1725\right),\ p\geq 1\}$   of $L^{2}([0, 1725]).$
    \item[Step 7.] The projections  in  Step 6 are spatially normalized.
\item[Step 8.]   The two--dimensional Fast Fourier Transform  is applied to the outputs of Step 7.
\item[Step 9.]  The empirical loss function  $\sigma_{400}(\boldsymbol{\theta})$   in  (\ref{wfs}) is computed from  Step 8.
 \item[Step 10.] Model (\ref{eqsdosarh1}) is fitted by computing  the minimum of the  constrained nonlinear multivariate function  $\sigma_{400}(\boldsymbol{\theta}).$ Here,  $L_{3}$   is not necessarily given by the composition of operators $L_{i},$ $ i=1,2,$ as in the previous section. We have used the option 'lbfgs' of \emph{fmincon}   MatLab function, which is an  optimization algorithm in the family of quasi-Newton methods, involving an inverse Hessian matrix  estimate   to steer its search through variable space.
     \item[Step 11.]  The   SAR$\mathcal{H}$(1) plug-in parametric predictor  (\ref{SARHpp}) is  obtained from Step 10.
          \end{itemize}

\begin{figure}
\begin{center}
\includegraphics[width=12cm,height=6cm]{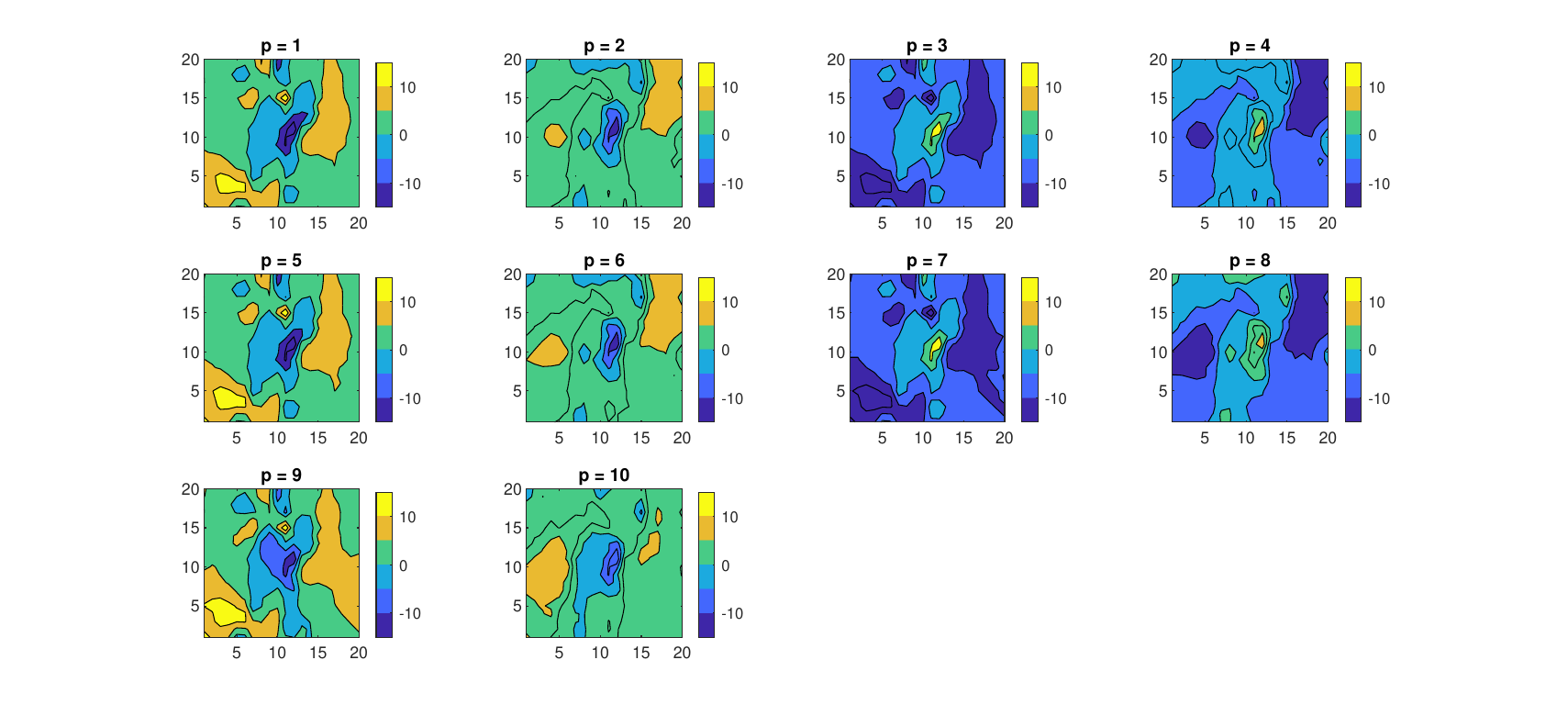}
\end{center}
\caption{Ten normalized  projections of the    spatial functional residual log--intensity}
\label{fig7}
\end{figure}

\begin{figure}
\begin{center}
\includegraphics[width=12cm,height=6cm]{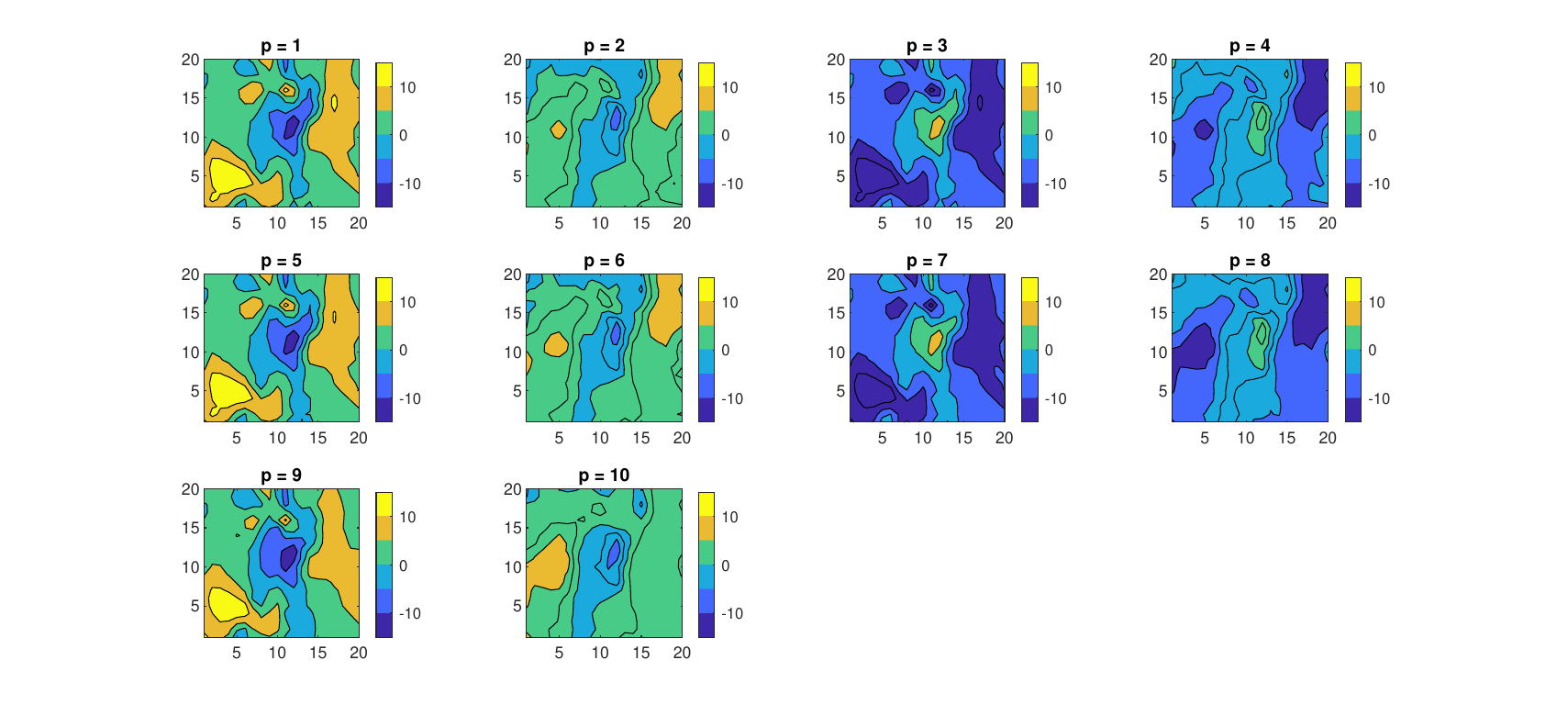}
\end{center}
\caption{Estimates of the normalized  projections of the  spatial     functional  residual log-intensity}
\label{fig8}
\end{figure}

Our parametric model fitting has been  performed in terms of the spectral kernel defined by  the eigenvectors $\{\sin\left(\pi p t/1725\right), \ p\geq 1\}$ of the Dirichlet negative Laplacian operator  on  $L^{2}([0, 1725]).$  The least-squares estimate of  Cox process values are obtained from Step 11  applying   (\ref{lspredictor})   in terms of $\{\sin\left(\pi p t/1725\right), \ p\geq 1\},$  and the corresponding inverse projection formula.
Figures \ref{fig7}--\ref{fig8} respectively show the original and estimated spatial  projections  of the functional residual log--intensity random field, for the truncation parameter $M=10$ (see outputs from Steps 6--7 and 10).

\begin{figure}
\begin{center}
\includegraphics[width=10cm,height=4.5cm]{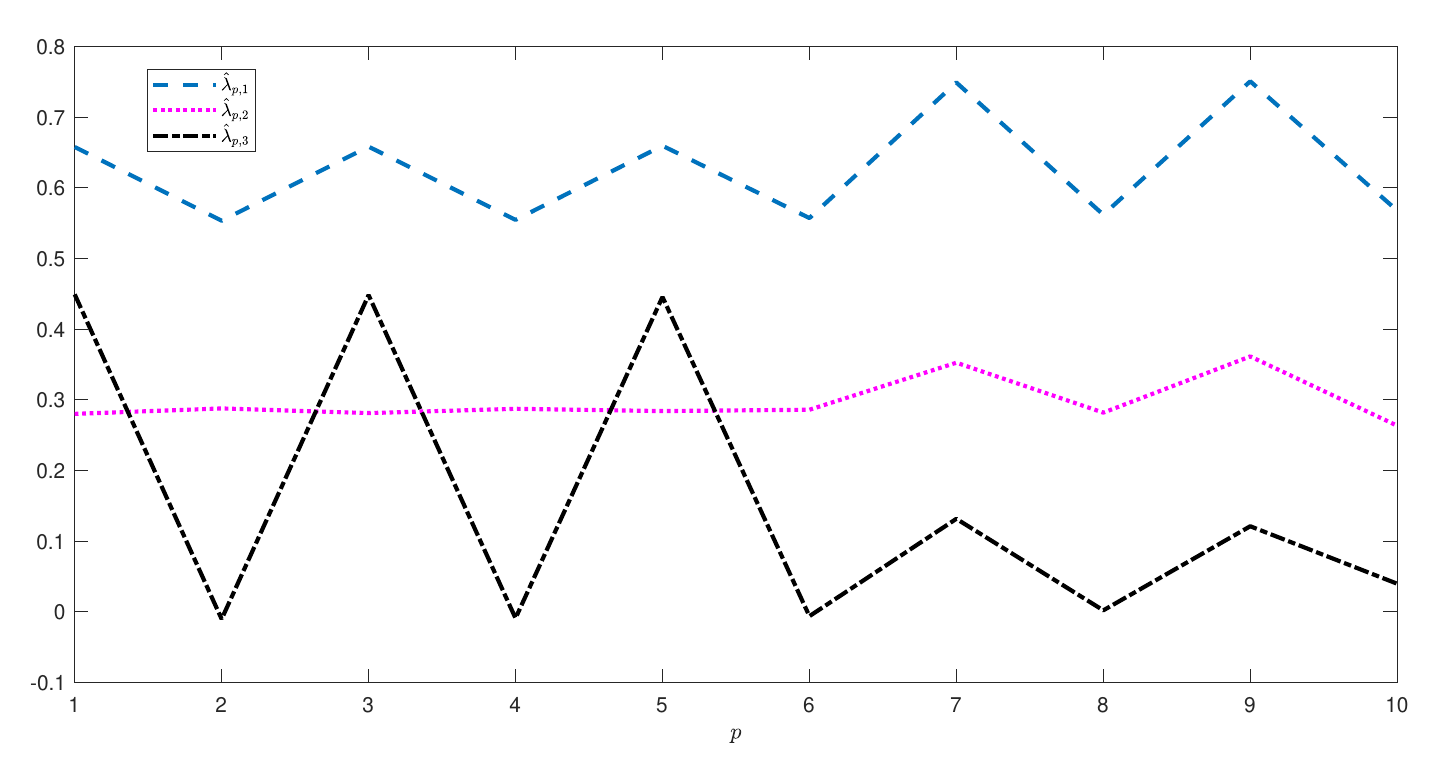}
\end{center}
\caption{Plot of point spectra estimates of $L_{i,\widehat{\boldsymbol{\theta}}_{i,\mathcal{N}}},$ $i=1,2,3,$ for $M=10$ and  $\mathcal{N}=400.$ In the plot,  we have used the notation $\widehat{\lambda}_{p,i}=\lambda_{p,i}(\widehat{\boldsymbol{\theta}}_{i,\mathcal{N}}),$  for $p=1,\ldots,10,$  $i=1,2,3$}
\label{fig13}
\end{figure}

\begin{figure}[h]
\begin{center}
\includegraphics[width=12cm,height=6cm]{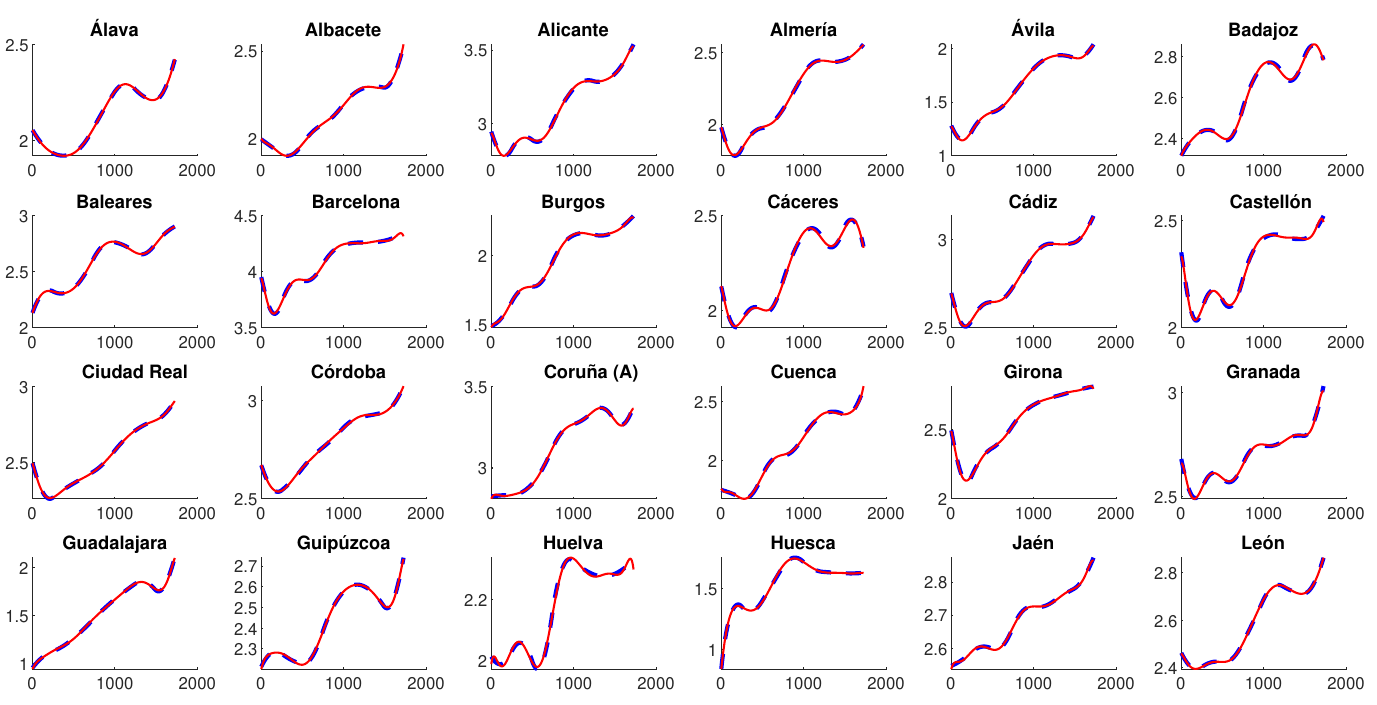}
\end{center}
\caption{Original respiratory--disease--mortality log--intensity curves (dashed blue line), and their respective estimates  (red line), for the above $24$ Spanish provinces displayed}
\label{fig5da}
\end{figure}

\begin{figure}[h]
\begin{center}
\includegraphics[width=12cm,height=6cm]{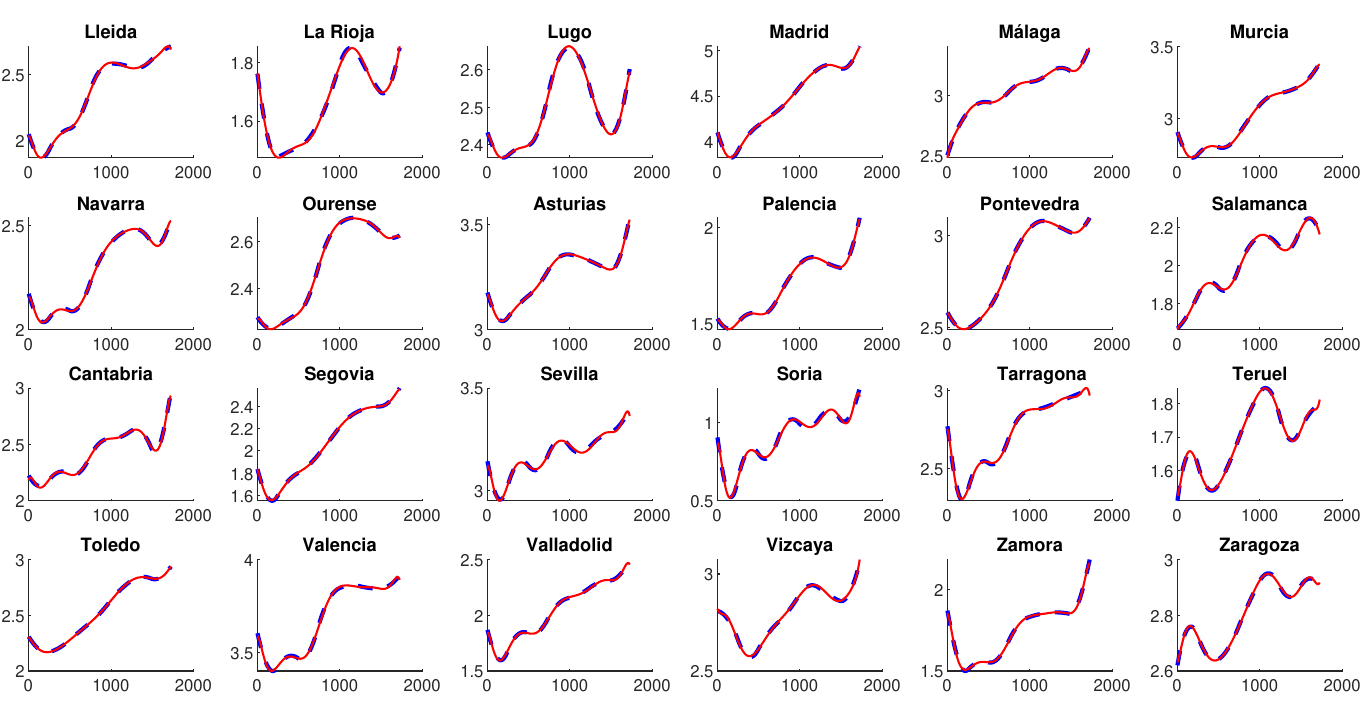}
\end{center}
\caption{Original respiratory--disease--mortality log--intensity curves (dashed blue line), and their respective estimates  (red line), for the above $24$ Spanish provinces displayed}
\label{fig5db}
\end{figure}

\begin{figure}[h]
\begin{center}
\includegraphics[width=10cm,height=4cm]{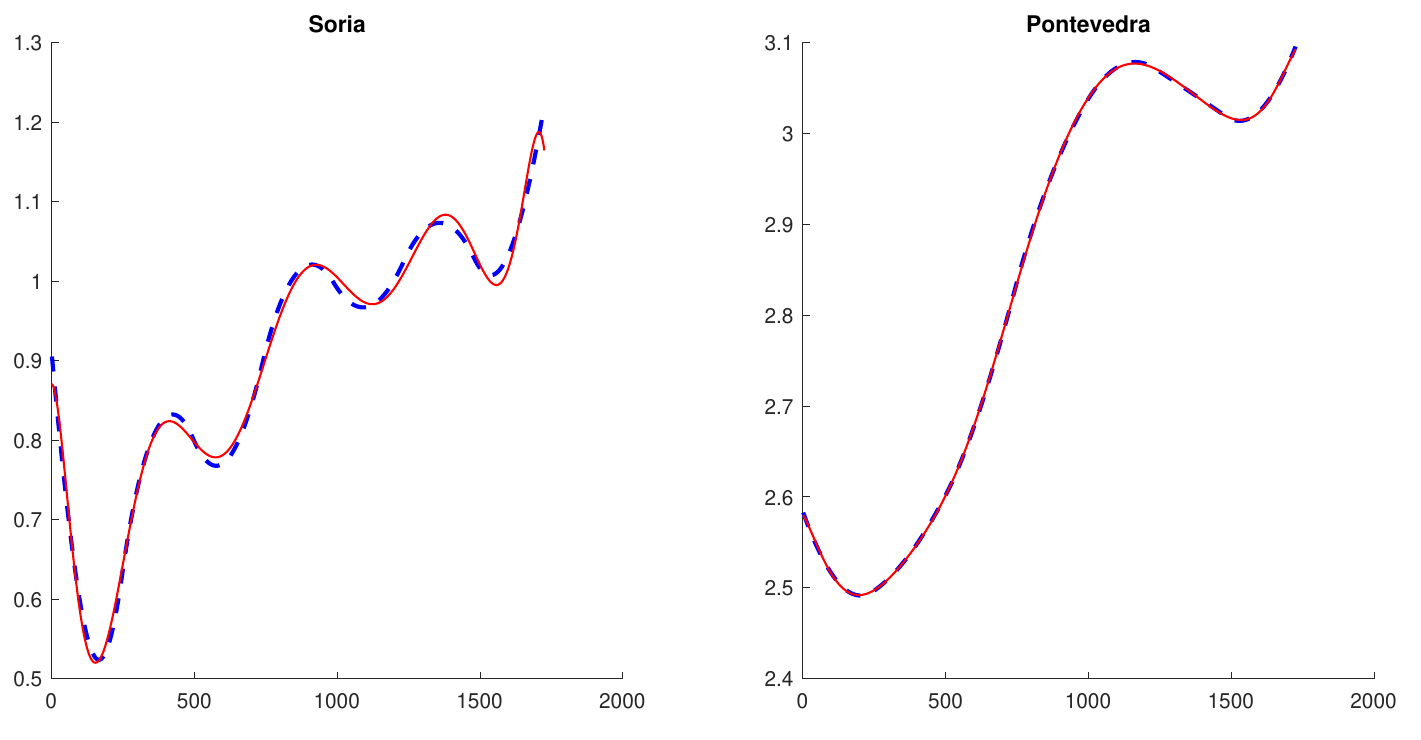}
\end{center}
\caption{Spanish provinces with largest, Soria  (left), and smallest, Pontevedra  (right), $L^{1}([0,1725])$--norm of the associated functional  absolute  relative  error are shown. Original curves  are plotted in dashed blue line, and estimated curves are represented in red line.}
\label{fig5d3}
\end{figure}

\begin{figure}
\begin{center}
\includegraphics[width=5.5cm,height=5.5cm]{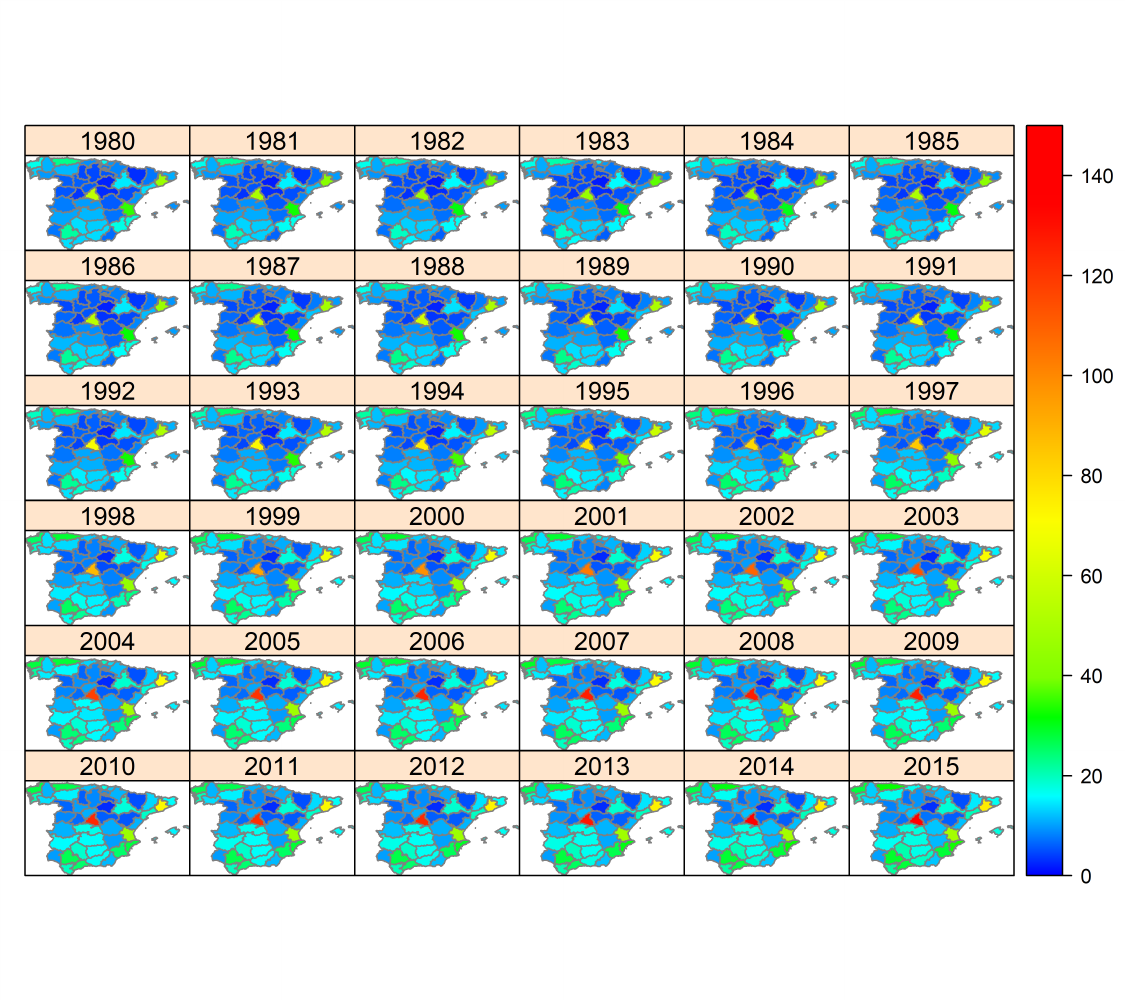}
\includegraphics[width=5.5cm,height=5.5cm]{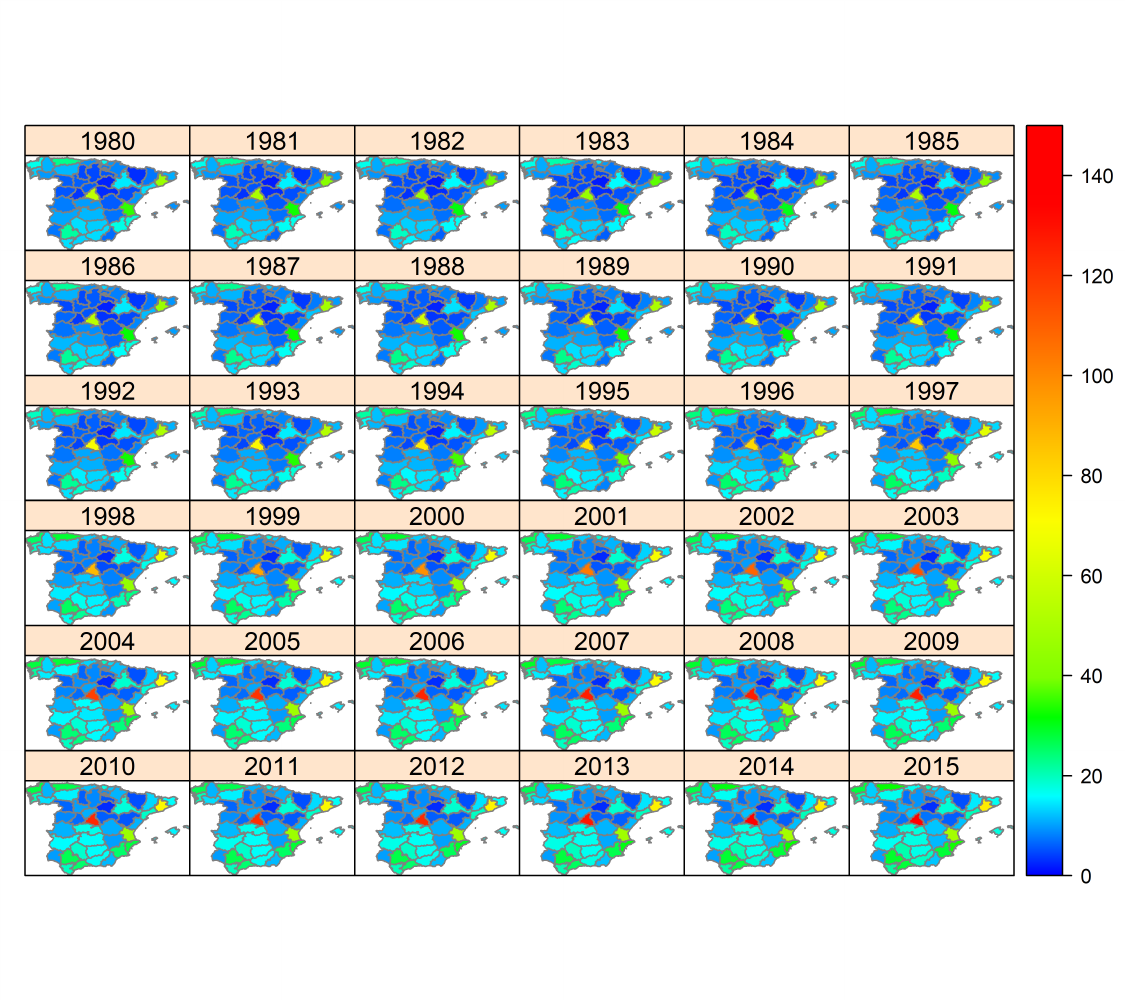}
\end{center}
\caption{Observed (left)  and estimated (right) annually averaged  respiratory--disease--mortality risk maps, during the period 1980--2015}\label{fig2}
\end{figure}

In Steps 9--10,  we have fitted  the following   parametric model
\begin{equation}\widehat{\lambda}_{p,i}=\lambda_{p,i}(\widehat{\boldsymbol{\theta}}_{i,\mathcal{N}})= \widehat{\theta}_{i,1,\mathcal{N}}+|\sin (p\pi/2)|\widehat{\theta}_{i,2,\mathcal{N}}(p),\ p=1,\dots, 10,\  i=1,2,3,\label{pmf}\end{equation}
\noindent for the eigenvalues of  $L_{i, \widehat{\boldsymbol{\theta}}_{i,\mathcal{N}}},$ $i=1,2,3,$  considering  $M=10,$ from a functional sample size $\mathcal{N}=400.$ The fitted parameter values  $\widehat{\boldsymbol{\theta}}_{i,\mathcal{N}},$ $i=1,2,3,$ are the following:
\begin{eqnarray}&&
\widehat{\theta}_{1,1,\mathcal{N}}
= 0.56; \  \widehat{\theta}_{1,2,\mathcal{N}}(p)=0,\  p=2,4,6,8,10; \nonumber\\
 && \widehat{\theta}_{1,2,\mathcal{N}}(p)= 0.08,\ p=1,3,5; \    \widehat{\theta}_{1,2,\mathcal{N}}(p)= 0.189;\ p=7,9; \nonumber\\
&& \widehat{\theta}_{2,1,\mathcal{N}}= 0.28;\  \widehat{\theta}_{2,2,\mathcal{N}}(p)= 0,\  p=1,2,3, 4,5, 6,8,10;\nonumber\\
&&\widehat{\theta}_{2,2,\mathcal{N}}(7)=0.0725; \  \widehat{\theta}_{2,2,\mathcal{N}}(9)= 0.0814;\
\widehat{\theta}_{3,1,\mathcal{N}}= 0.0033; \nonumber\\&&  \widehat{\theta}_{3,2,\mathcal{N}}(p)=-0,0052, \  p=2,4,6;\   \widehat{\theta}_{3,2,\mathcal{N}}(p)=0,\  p=8,10; \nonumber\\
&& \widehat{\theta}_{3,2,\mathcal{N}}(p)= 0.4444, \ p=1,3,5; \
\widehat{\theta}_{3,2,\mathcal{N}}(p)= 0,1230, \ p=7,9,\quad  \mathcal{N}=400.\nonumber\\\label{pvfitted}
\end{eqnarray}
The  corresponding parametric  estimations $\left\{\lambda_{p,i}(\widehat{\boldsymbol{\theta}}_{i,\mathcal{N}}), \ p=1,\dots,10\right\},\ i=1,2,3,$ of the point spectra of  $L_{i,\widehat{\boldsymbol{\theta}}_{i,\mathcal{N}}},$ $i=1,2,3,$  are displayed in Table \ref{tbste10} and Figure  \ref{fig13}. Finally,  the original respiratory--disease--mortality log-intensity curves (dashed blue line), and their estimations   (red line), at the  $48$ Spanish provinces, are displayed in Figures \ref{fig5da}--\ref{fig5db}. The  worst  (Soria province) and the best (Pontevedra province) model fitting are  zoomed in
Figure \ref{fig5d3}. Also, the original and estimated  annually averaged respiratory--disease--mortality  risk maps   are  displayed  in  Figure \ref{fig2}.
\begin{table}
\begin{center}
\begin{tabular}{cccc}

\hline

	&	$\lambda_{p,1}(\widehat{\boldsymbol{\theta}}_{1,\mathcal{N})}$&$\lambda_{p,2}(\widehat{\boldsymbol{\theta}}_{2,\mathcal{N}})$&$
\lambda_{p,3}(\widehat{\boldsymbol{\theta}}_{3,\mathcal{N}})$\\\hline

 $p=1$&   0.6578&    0.2801   & 0.4493\\

$p=2$&    0.5534&    0.2878&   -0.0102\\

$p=3$&    0.6583&    0.2812&    0.4482\\

$p=4$&    0.5547&    0.2873&   -0.0091\\

$p=5$&    0.6595 &   0.2841&    0.4456\\

$p=6$&    0.5572 &   0.2860&   -0.0061\\

$p=7$&    0.7490&    0.3525&    0.1316\\

$p=8$&    0.5620&    0.2819&    0.0021\\

$p=9$&    0.7507&    0.3614&    0.1211\\

$p=10$&    0.5678&    0.2634 &   0.0398\\

\hline

\end{tabular}
\end{center}
\caption{Point spectra estimates of $L_{i, \widehat{\boldsymbol{\theta}}_{i,\mathcal{N}}},$ $i=1,2,3,$ for $M=10,$  $\mathcal{N}=400$}
\label{tbste10}
\end{table}


\medskip

To assess  spatial functional prediction ability of our approach,   cross-\linebreak validation is implemented.   By leaving aside  the curves observed at the nodes  in a neighborhood of the province defining the region of interest (the validation functional data set), Steps 1--11 are run in terms of the  remaining functional observations, spatially distributed at the neighborhoods of the rest of the  Spanish provinces   (the  training functional data set). The corresponding SAR$\mathcal{H}$(1)  plug-in parametric predictor  (\ref{SARHpp}) is computed at each iteration. After  $48$ iterations,   the   mean cross-validation  functional   absolute relative error $\mbox{CVFARE} (t) = \frac{1}{48}\sum_{\mathbf{z}\in \mathcal{R}}\left|\frac{\varkappa_{\mathbf{z}}(t)-\widehat{\varkappa }_{\mathbf{z}}(t)}{\varkappa_{\mathbf{z}}(t)}\right|,$ $t=1,\dots, 1725,$ is obtained,  where $\mathcal{R}$ denotes the set of latitudes and longitudes defining the spatial locations of the $48$  Spanish provinces.
Table \ref{LOOCVy} displays the annual   mean of the pointwise values of CVFARE.
The empirical value   $\|\mbox{CVFARE}\|_{L^{1}\left([0,1725]\right)}=0.0016$ computed from  Table \ref{LOOCVy} shows    a good spatial functional predictive ability of the   presented approach  beyond the log--Gaussian scenario.
\begin{table}
\caption{ACVFARE. Pointwise annually averaged CVFARE obtained from the  validations}
\label{LOOCVy}
\begin{tabular}{cccccccc}
\hline
Year& ACVFARE &Year& ACVFARE &Year& ACVFARE &Year& ACVFARE\\\hline
1980&0.00301 & 1989&0.00322  &1998&0.00135&2007&0.00007\\
1981&0.00273 & 1990&0.00245  &1999&0.00036&2008&0.00008\\
1982&0.00137 & 1991&0.00073  &2000&0.00106&2009&0.00008\\
1983&0.00295 & 1992&0.00116  &2001&0.00191&2010&0.00053\\
1984&0.00139 & 1993&0.00315  &2002&0.00134&2011&0.00124\\
1985&0.00192 & 1994&0.00357  &2003&0.00033&2012&0.00092\\
1986&0.00323 & 1995&0.00129  &2004&0.00085&2013&0.00159\\
1987&0.00175 & 1996&0.00276  &2005&0.00100&2014&0.00142\\
1988&0.00118 & 1997&0.00244  &2006&0.00045&2015&0.00230\\\hline
\end{tabular}
\end{table}

\section{Concluding remarks}
\label{s7}
 The results derived in this paper can be easily  reformulated for $\mathcal{H}$--valued spatially continuous log--intensity models, replacing  the index set $[-\pi,\pi]^{d}$ by $\mathbb{R}^{d},$ and adapting  the subsequent conditions to an unbounded frequency domain,  in the definition of  our spectral density operator family.   Indeed, one can adopt our discrete spatial parametric framework for practical purposes, considering  constant functional values over  the spatial  regular grid quadrants (\cite{Rathbun94}). As usual, for the  approximation of a continuous log--intensity model, interpolation and smoothing techniques can  be implemented. For example, spatial smoothing based on spline bases has been widely applied during the last decades, since the eighties (see, e.g.,   \cite{Ogata88}).

  Several problems remain open in the field of FDA techniques applied to the statistical analysis of spatial point patterns
  (see, e.g.,   \cite{HorKok12}; \cite{Ruiz12};
  \cite{GB2016}; \cite{Daimuller2018}, among others).
 This paper provides a spatial functional spectral--based approach,  in the parametric estimation of the second order moments of the   Cox process family, driven  by a spatial  $\mathcal{H}$--valued  log--intensity. The corresponding functional analysis of variance can also be achieved from the derived parametric predictor.
       One of the key  problems to be faced, in a near future, is the definition of a multivariate version of the introduced  point process  class, as well as its  spectral functional estimation in an infinite--dimensional multivariate framework. This subject  constitutes the topic of a subsequent paper.

\section*{Appendix. Proof of Theorem \ref{th1ex}}
In the derivation of the proof of Theorem \ref{th1ex},  the truncated empirical expansion of the periodogram operator
\begin{eqnarray}&&
\mathcal{I}_{\boldsymbol{\xi }}^{\mathcal{N}}(\varphi_{k})(\varphi_{l})=\widetilde{X}^{\mathcal{N}}_{\boldsymbol{\xi }
}(\varphi_{k}) \widetilde{X}^{\mathcal{N}}_{-\boldsymbol{\xi } }
(\varphi_{l})\nonumber\\ &&=\frac{1}{(2\pi)^{d}}\sum_{\mathbf{z}\in \prod_{i=1}^{d}[-\mathcal{N}_{i}+1,\mathcal{N}_{i}-1]}
\mathcal{C}(\mathbf{z},k,l)\exp\left(-i\sum_{j=1}^{d}\xi_{j}z_{j}\right),\quad k,l\geq 1\nonumber\\
&&\hspace*{1cm} \mathbf{z}=(z_{1},\dots, z_{d})\in \mathbb{Z}^{d},\quad z_{i}\in [-\mathcal{N}_{i}+1,\mathcal{N}_{i}-1],\ i=1,\dots,d, \label{fepo}\end{eqnarray} \noindent in terms of the Fourier coefficients introduced  in (\ref{deffcro}) is first applied. At the same time, the truncated Fourier series  (\ref{eqccsis})  defined from the Fourier coefficients
\begin{eqnarray}
g(\mathbf{z},\theta, k)&=&\frac{1}{(2\pi)^{d}}\int_{[-\pi,\pi]^{d}}\exp\left(i\sum_{j=1}^{d}\varpi_{j}z_{j}\right)\mathcal{F}_{\boldsymbol{\varpi},\theta}^{-1}(\varphi_{k})(\varphi_{k})
d\boldsymbol{\varpi}\nonumber\\
&& \hspace*{2cm}  \mathbf{z}=(z_{1},\dots,z_{d})\in\mathbb{Z}^{d},\quad \theta \in \Theta,\quad  k\geq 1,\nonumber\\\label{FCE}
\end{eqnarray}
\noindent   of the inverse $\mathcal{F}^{-1}_{\boldsymbol{\varpi},\theta }$ of the spectral density operator is also considered.
 Equations (\ref{fepo})--(\ref{FCE}) are applied in the derivation of the convergence to zero of equations (\ref{eqlimbb}) and (\ref{ftthpe}).
Specifically, let $\xi_{z_{j}}=\frac{2\pi z_{j}}{\mathcal{N}_{j}},$ with  $z_{j}\in \mathbb{Z}$ such that   $-\frac{\mathcal{N}_{j}}{2}< z_{j}\leq
\left[\frac{\mathcal{N}_{j}}{2}\right],$ for $j=1,\dots,d,$ and  $\mathcal{N}=\prod_{j=1}^{d}\mathcal{N}_{j}$ being  the
 functional sample  size.
   Applying  Parserval identity in $L^{2}([-\pi,\pi]^{d}),$ for  $M$ fixed  large, and  $\theta \in \Theta,$
\begin{eqnarray} &&
 \frac{(2\pi)^{d}}{\mathcal{N}}\sum_{\mathbf{z}_{j}}
 \left|q^{M}_{\boldsymbol{\xi}_{\mathbf{z}_{j}},\theta}\mathcal{I}_{\boldsymbol{\xi}_{\mathbf{z}_{j}}}^{(\mathcal{N})}(\varphi_{k})(\varphi_{k})
 -q^{M}_{\boldsymbol{\xi}_{\mathbf{z}_{j}},\theta}
 \mathcal{F}_{\boldsymbol{\xi}_{\mathbf{z}_{j}} ,\theta_{0} }(\varphi_{k})(\varphi_{k})\right|^{2}\nonumber\\
&&=\frac{(2\pi)^{d}}{\mathcal{N}} \sum_{u_{j}\in [-M_{j}+1, M_{j}-1],\ j=1,\dots,d}\left|\prod_{j=1}^{d}\left(1-\frac{|u_{j}|}{M_{j}}\right)g(\mathbf{u},\theta ,k)\right|^{2}
\nonumber\\ &&\hspace*{3cm}\times \left| [\mathcal{C}(\mathbf{u},k,k)+\mathcal{K}(\mathcal{N})] - \mathcal{R}_{\mathbf{u},\theta_{0}}(\varphi_{k})(\varphi_{k})\right|^{2},
  \label{czeroas}
  \end{eqnarray}
  \noindent where $\mathcal{K}(\mathcal{N})\to 0,$ as $\mathcal{N}\to \infty$  (see \cite{Hannan73}).  \textbf{Assumption A2} then  implies the a.s. convergence to zero of (\ref{czeroas}),  as $\mathcal{N}\to \infty,$  uniformly in $k\geq 1,$ in view of equation (\ref{fcshnormas}).

 We now proceed to detail the derivation  of the  strong--consistency of $\widehat{\theta}_{\mathcal{N}},$
from equations  (\ref{limas}) and (\ref{eqineqhh}). Specifically,  if, as $\mathcal{N}\to \infty,$ $\widehat{\theta}_{\mathcal{N}}$ does not converge a.s. to
 $\theta_{0},$ there exists a subsequent $\widehat{\theta}_{\mathcal{N}_{m}}$ converging  to $\theta^{\prime}\neq \theta_{0},$ as $m\to \infty$ ($\mathcal{N}_{m}\to \infty$),  such that
 $\theta^{\prime}\in \Theta.$   From (\ref{limas}),  for a given positive constant  $\eta >0,$
 the  a.s. lower limit  $\underline{\lim}_{\mathcal{N}_{m}\to \infty}\sigma_{\mathcal{N}_{m}}(\widehat{\theta}_{\mathcal{N}_{m}})$ satisfies
\begin{eqnarray}
&&\hspace*{-0.3cm}\underline{\lim}_{\mathcal{N}_{m}\to \infty}\sigma_{\mathcal{N}_{m}}(\widehat{\theta}_{\mathcal{N}_{m}})\geq \sup_{k\geq 1}\frac{1}{[2\pi]^{d}}\int_{[-\pi,\pi]^{d}}\mathcal{F}_{\boldsymbol{\xi},\theta_{0}}\left[\mathcal{F}_{\boldsymbol{\xi},\theta^{\prime }}+ \eta I_{\mathcal{H}+i\mathcal{H}}\right]^{-1}(\varphi_{k})(\varphi_{k}) d\boldsymbol{\xi},
\nonumber\end{eqnarray}
\noindent where   $I_{\mathcal{H}+i\mathcal{H}} $ denotes the identity operator on $\mathcal{H}+i\mathcal{H}.$
From equation (\ref{eqineqhh}), for $\eta $ sufficiently small, $\underline{\lim}_{\mathcal{N}_{m}\to \infty}\sigma_{\mathcal{N}_{m}}(\widehat{\theta}_{\mathcal{N}_{m}})> 1,$ a.s.
On the other hand, for every $\theta\in \Theta,$   $\overline{\lim}_{\mathcal{N}_{m}\to \infty}\sigma_{\mathcal{N}_{m}}(\widehat{\theta}_{\mathcal{N}_{m}})\leq \overline{\lim}_{\mathcal{N}_{m}\to \infty}\sigma_{\mathcal{N}_{m}}(\theta ),$ a.s. Hence, for every $\theta\in \Theta,$
\begin{eqnarray}&&\overline{\lim}_{\mathcal{N}_{m}\to \infty}\sigma_{\mathcal{N}_{m}}(\widehat{\theta}_{\mathcal{N}_{m}})\leq
\sup_{k\geq
1}\frac{1}{[2\pi]^{d}}\int_{[-\pi,\pi]^{d}}\mathcal{F}_{\boldsymbol{\xi},\theta_{0}}\mathcal{F}_{\boldsymbol{\xi},\theta}^{-1}(\varphi_{k})(\varphi_{k})
d\boldsymbol{\xi }\nonumber,\label{eqineqbbc2}
\end{eqnarray} \noindent which implies  the  \mbox{a.s.} inequality
$$\overline{\lim}_{\mathcal{N}_{m}\to \infty}\sigma_{\mathcal{N}_{m}}(\widehat{\theta}_{\mathcal{N}_{m}})\leq \inf_{\theta\in \Theta }\sup_{k\geq 1}\frac{1}{[2\pi]^{d}}\int_{[-\pi,\pi]^{d}}\mathcal{F}_{\boldsymbol{\xi},\theta_{0}}\mathcal{F}_{\boldsymbol{\xi},\theta}^{-1}(\varphi_{k})
(\varphi_{k})d\boldsymbol{\xi}= 1,$$\noindent  in view of   (\ref{infbb}), leading to
a contradiction. Therefore, $\theta^{\prime}=\theta_{0}=\lim_{\mathcal{N}\to
\infty}\widehat{\theta}_{\mathcal{N}},$ a.s., as we wanted to prove.

\section*{Acknowledgements}
This work has been supported in part by projects
PGC2018-099549-B-I00,  MTM2016-78917-R,  and PID2019-107392RB-100 of the Ministerio de Ciencia, Innovaci\'on y Universidades, Spain (co-funded with FEDER funds), and ERDF Operational Programme 2014--2020 and the Economy and Knowledge
Council of the Regional Government of
Andalusia, Spain (A-FQM-345-UGR18).

\end{document}